\documentclass{iopjournal_preprint}
\usepackage{amsmath,amssymb,amsfonts}
\usepackage{algorithmic}
\usepackage{graphicx}
\usepackage{textcomp}
\usepackage{tikz}
\usepackage{hyperref}
\usepackage{bm}
\usepackage{multirow}
\usepackage{subcaption}
\usepackage[export]{adjustbox}
\usepackage{booktabs}
\usepackage{pdfpages}

\usepackage{xparse}

\usepackage{expl3}
\ExplSyntaxOn
\cs_if_exist:NF \tl_length:N
{
	\cs_new_eq:NN \tl_length:N \tl_count:N
}
\ExplSyntaxOff

\DeclareMathOperator*{\argmin}{argmin}
\usepackage[T1]{fontenc}
\usepackage[backend=biber, style=ieee, url=false]{biblatex}
\AtEveryBibitem{\clearfield{urldate}}
\renewcommand{\cite}{\parencite}

\newcommand{\ArticleTitle}{Real-time 3D Ultrasonic Needle Tracking with a Photoacoustic Beacon}
\newcommand{\KingsAddress}{School of Biomedical Engineering and Imaging Sciences, King's College London, London, SE1 7EU, UK.}
\newcommand{\GSTTAddress}{Guys and St Thomas' NHS Trust, London, SE1 7EH, UK.}
\newcommand{\UCLHAddress}{University College London Hospitals NHS Foundation Trust, London, NW1 2BU, UK.}
\newcommand{\BICIAddress}{GBA Institute of Collaborative Innovation, Guangzhou, China.}
\newcommand{\UCLAddress}{Medical Physics and Biomedical Engineering, University College London, London, WC1E 6BT, UK.}

\newcommand{\AuthorORCIDA}{0000-0001-6200-9201}

\newcommand{\AuthorORCIDB}{0000-0002-2528-4818}

\newcommand{\AuthorORCIDC}{0000-0001-8657-0546}

\newcommand{\AuthorORCIDD}{0000-0002-1969-0035}

\newcommand{\AuthorORCIDE}{0009-0001-9975-8537}

\newcommand{\AuthorORCIDF}{0000-0002-5694-5340}

\newcommand{\AuthorORCIDG}{0000-0002-8366-4886}

\newcommand{\AuthorORCIDH}{0000-0001-9970-0522}

\newcommand{\AuthorORCIDI}{0000-0002-1932-1811}

\newcommand{\AuthorORCIDJ}{0000-0002-1147-6886}

\makeatletter
\def\CreateAuthorFullName#1{%
	\expandafter\newcommand\csname AuthorFullName#1\endcsname{%
		\csname AuthorFirstName#1\endcsname\ \csname AuthorLastName#1\endcsname%
	}%
}
\@for\i:=A,B,C,D,E,F,G,H,I,J\do{%
	\expandafter\CreateAuthorFullName\expandafter{\i}%
}
\makeatother

\begin{document}
	
\includepdf[pages={1}]{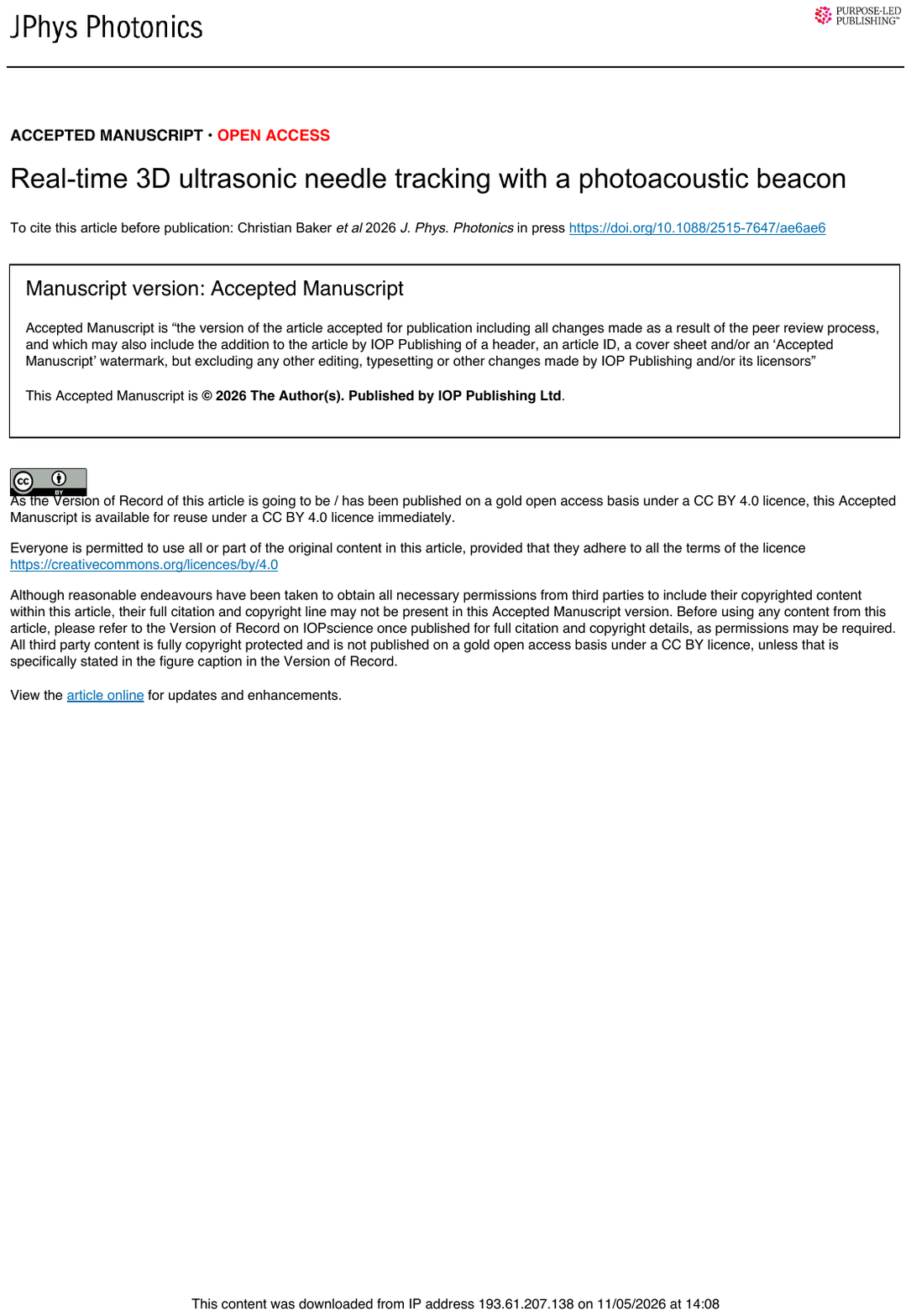}

\articletype{Paper}

\title{\ArticleTitle}

\author{
	\AuthorFullNameA$^{1,*}$\orcid{\AuthorORCIDA},  
	\AuthorFullNameB$^1$\orcid{\AuthorORCIDB}, 
	\AuthorFullNameC$^2$\orcid{\AuthorORCIDC},
	\AuthorFullNameD$^3$\orcid{\AuthorORCIDD}, 
	\AuthorFullNameE$^1$\orcid{\AuthorORCIDE},
	\AuthorFullNameF$^1$\orcid{\AuthorORCIDF},
	\AuthorFullNameG$^4$\orcid{\AuthorORCIDG},
	\AuthorFullNameI$^2$\orcid{\AuthorORCIDH},
	\AuthorFullNameH$^5$\orcid{\AuthorORCIDI},
	\AuthorFullNameJ$^{1,\dagger}$\orcid{\AuthorORCIDJ}  
}

\affil{$^1$\KingsAddress}

\affil{$^2$\UCLAddress}

\affil{$^3$\BICIAddress}

\affil{$^4$\UCLHAddress}

\affil{$^5$\GSTTAddress}

\affil{$^*$Primary corresponding author.}

\affil{$^\dagger$Secondary corresponding author.}

\email{christian.baker@kcl.ac.uk, wenfeng.xia@kcl.ac.uk}


\begin{abstract}
Many minimally invasive procedures, such as core needle biopsy of focal liver lesions, nerve blocks, and fetal and vascular interventions, are typically performed under ultrasound guidance, which provides real-time, high-resolution visualisation of tissue anatomy. Accurate and efficient localisation of the needle tip relative to patient anatomy is essential for guiding the needle towards the procedure target, avoiding adverse events and reducing the need for repeat procedures. However, the 3D nature of the procedure and poor image contrast of the needle in heterogeneous tissue or at steep insertion angles often lead to confusion over the true location of the tip within the 2D guidance images, and existing methods to enhance needle visibility largely remain limited to 2D. Here, we present a novel interventional ultrasound system capable of 2D B-mode imaging and 3D needle tracking. The tip location is determined from the time-of-flight of ultrasound generated by a photoacoustic beacon embedded in the needle bevel and received by a sparse receiver array distributed around the imaging system’s curvilinear ultrasound probe. The measured tracking accuracy was better than 2\,mm for depths up to 140\,mm in water, and approximately 2\,mm on average in an \textit{ex vivo} tissue phantom, with referenced positions derived from X-ray computed tomography. In a usability study involving 12 clinicians performing biopsy procedures in a \textit{ex vivo} tissue phantom, the failure rate was reduced by 35\%, from 15.8\% to 10.3\% after only a few minutes of training. These results demonstrate that the proposed system has strong potential to support a wide range of minimally invasive procedures by enabling clinicians to accurately target anatomical structures with millimetre-level precision, improving the efficiency and effectiveness of diagnostic sampling and therapeutic delivery or ablation, and reducing the risk of adverse events.
\end{abstract}

\section{Introduction}
\label{sec:introduction}
Ultrasound is widely used to guide minimally invasive procedures across multiple clinical disciplines including oncology, regional anaesthesia and pain management, and fetal medicine, owing to its real-time visualisation of tissue anatomy, broad accessibility, and cost-effectiveness. For example, in focal liver lesion procedures, such as biopsies, ablation and fiducial marker placement, precise needle tip localisation relative to patient anatomy is critical for directing the needle to the intended target, minimising adverse events, and reducing the need for repeat procedures.
The depths of the targets (typically up to 10\,cm and sometimes beyond \cite{zengFocalLiverLesions2024}) and proximity of critical structures require careful 3D manoeuvring of the needle within the patient over several minutes.
Adjustments should ideally be made while the needle is still in shallow tissue, limiting punctures to and manoeuvres within the organ to reduce the risk of pain, bleeding, and trauma.
Once the needle penetrates the liver, accurate localisation of the needle tip is required to reach the target lesion, which can be as small as 5 mm.

Accurate 3D needle-tip localisation is challenging using the 2D ultrasound imaging with which these procedures are typically guided.
The high heterogeneity of shallow subcutaneous adipose tissue results in poor image contrast, which can make it exceedingly difficult to locate the needle tip.
At steep insertion angles, specular reflections from the needle are directed away from the imaging probe, further reducing contrast.
Due to the inherently 3D nature of the procedures, the needle shaft and the ultrasound imaging plane are often misaligned---intentionally or otherwise---which can lead to misinterpretation of the needle shaft’s cross-section as its tip.

Approximately 30\% of liver biopsy procedures need to be repeated due to inadequate sampling \cite{vernuccioNegativeBiopsyFocal2019a}, with a 7\% false-negative rate \cite{khalifaUtilityLiverBiopsy2020a}.
Up to 5\% of procedures result in significant bleeding, and 1\% require hospitalisation \cite{khalifaUtilityLiverBiopsy2020a}.
Outcomes are highly dependent on the clinicians’ skills, and senior staff are often required to undertake more challenging procedures.
It is also common for X-ray computed tomography (CT) to be used to guide the most difficult insertions, or where anatomy restricts visualisation of the needle with 2D ultrasound.
CT is time-consuming, costly, and exposes patients and clinicians to ionising radiation.
The need to repeatedly remove patients from the scanner so that the needle can be advanced adds to procedural inefficiency, as CT cannot provide real-time imaging.

The challenges inherent to ultrasound guided percutaneous needle procedures have led to the development of tools such as needle guides and echogenic needles.
Needle guides, which attach to an ultrasound probe to maintain alignment of the needle with the imaging plane, are inappropriate due to the need for frequent 3D manoeuvres requiring repositioning of the needle and probe.
Echogenic needles can enhance visibility of the needle within the 2D ultrasound image, particularly at steep insertion angles, but are not visible outside of the imaging plane.

Active needle tip tracking---where signals are detected or generated by transducers integrated into the needle tip---has the potential to provide clinicians with accurate real-time knowledge of the location of the tip even when obscured in the guidance image.
Investigated technologies include: electromagnetic (EM) sensors \cite{hakimeElectromagneticTrackedBiopsyUltrasound2012}; piezoelectric ultrasound sensors \cite{kasineNeedleTipTracking2019} and transmitters \cite{langbergEchotransponderElectrodeCatheter1988}; and fibre-optic ultrasound sensors \cite{xiaLookingImagingPlane2017, mathewsUltrasonicNeedleTracking2022, bakerIntraoperativeNeedleTip2022} and transmitters \cite{xiaUltrasonicNeedleTracking2017, ledijubellPhotoacousticbasedVisualServoing2018}.

For EM tracking, the needle tip is localised via detection of a known EM field.
This method can provide 3D localisation but is sensitive to EM disturbances from ferromagnetic material in the surgical environment and requires a bulky EM field generator close to the surgical site.
Tracking resolution can be worse than 3 mm \cite{beigiEnhancementNeedleVisualization2021} and the single-use needles can cost between \$100 and \$500 \cite{zhaoElectromagneticTrackingNeedle2019}.
Ultrasonic methods can provide sub-millimetre tracking accuracy, typically via an embedded piezoelectric transmitter or receiver.
Notably, Phillips / B.
Braun’s Onvision system incorporates an integrated piezoelectric receiver \cite{kasineNeedleTipTracking2019}.
However, piezoelectric transducers face inherent performance limitations, including high cost, and a necessary compromise between sensitivity, size, bandwidth and directivity.

Compared to piezoelectric solutions, fibre-optic ultrasound sensors and transmitters are more cost-effective and---being highly miniaturisable---are compatible with small needle gauges.
Their broadband capabilities accommodate a wide range of ultrasound bandwidths, and their omnidirectional nature reduces dependence on needle insertion angles.
Recent works have utilised miniature fibre-optic ultrasound sensors to detect transmissions from the ultrasound imaging array being used for guidance \cite{xiaLookingImagingPlane2017, mathewsUltrasonicNeedleTracking2022, bakerIntraoperativeNeedleTip2022}.
We previously demonstrated a 2D tracking system that works as an adjunct to a clinical ultrasound imaging system, using the imaging pulse sequence for tracking and therefore allowing truly simultaneous imaging and tracking \cite{bakerIntraoperativeNeedleTip2022}.

Tracking of a needle-integrated fibre-optic ultrasound transmitter was previously demonstrated using a research ultrasound system to acquire interleaved tracking and imaging frames from a linear array ultrasound transducer with a bandwidth of 5--14\,MHz \cite{xiaUltrasonicNeedleTracking2017}.
Fourier-domain time-reversal was used to reconstruct the location of the acoustic source from the signals received from the fibre-optic transmitter.
This interleaved method limited the acquisition frame rate of both imaging and the PA reconstruction was only possible in post-processing (the reconstruction step took approximately 0.5\,s per frame using a personal computer).
Due to the reliance on a 1D linear array, tracking was only possible in 2D, confined to the imaging plane.
Tracking accuracy was only assessed over relatively short tissue depths (up to 50\,mm).

Solutions that rely on a 1D linear or curvilinear array are unable to directly quantify the elevational (i.e., out-of-plane) location of the needle tip and therefore cannot quantitatively visualise the 3D location of the needle tip.
Attempting to account for this limitation, Tanaka et al.~\cite{tanakaSplitbasedElevationalLocalization2024} have proposed 3D tracking of a fibre-optic ultrasound transmitter through analysis of the shape of the received waveforms: positioning the transmitter out-of-plane introduced a time delay across the rectangular receiving elements, which was related to the distance of the needle tip from the imaging plane, allowing indirect quantification of elevational position.
Machine learning has also been used to estimate the distance of the needle tip from the imaging plane from such signals, but it was only tested in homogeneous media, and it was found that the signal-to-noise ratios were too poor for tracking further than approximately 6\,mm on each side of the imaging plane \cite{arjasNeuralNetworkKalman2022}.
Our previous work visualised the uncertainty in 2D tracked position by changing the colour, size or opacity of the tracking cursor, which was found to be useful for identifying if the needle was in-plane, but not quantifying its elevational position~\cite{bakerIntraoperativeNeedleTip2022}.
The accuracies of these approaches are likely to be spatially dependent and affected by the heterogeneity of tissues in and around the imaging plane, which will affect the morphologies of the received signals.

Quantitative 3D tracking has previously been achieved through the use of a research ultrasound system and custom 2D ultrasound array \cite{xiaLookingImagingPlane2017}.
Dedicated tracking elements aligned parallel to and on each side of a central linear imaging array transmitted Golay-coded ultrasound signals to a fibre-optic hydrophone integrated into the needle.
The lateral location of the needle tip was estimated from analysis of the amplitudes of the signals received by the tracking array, while localisation of the tip in the axial-elevational plane was achieved using multilateration (MLAT) of the times-of-arrival of the signals at the array.
While this method was able to provide a tracking accuracy better than 0.4\,mm at depths of up to 40\,mm and elevational locations up to 15\,mm from the imaging plane, clinical deployment was hindered by the requirement for custom ultrasound transmission sequences, performing imaging and tracking one after the other, limiting the frame rates of both and requiring a research-mode imaging system capable of exciting 256 channels (128 for imaging and 128 for tracking).
Furthermore, 40\,mm is too shallow for liver biopsy and other solid-organ interventions.
We have previously demonstrated 3D imaging and needle tracking using a sparse matrix imaging array and needle-integrated fibre-optic ultrasound sensor~\cite{liangVolumetricUltrasoundImaging2025}, but the depth was limited to 40\,mm.

This paper presents a novel interventional ultrasound system that provides simultaneous real-time ultrasound imaging and 3D ultrasonic needle tracking. The system utilises a trackable needle with an integrated fibre-optic ultrasound transmitter (photoacoustic beacon) to communicate with a sparse, 16-element receiver array attached to a clinical ultrasound imaging probe. An MLAT algorithm is employed for real-time determination of the location of the needle tip from the received signals, delivering precise 3D tracking beyond the capabilities of existing approaches. The tracking accuracy is presented, determined both in water using a 3-axis motion system and in an \textit{ex vivo} tissue phantom, in comparison to CT.
The results of a simulation study of the effects of sound speed heterogeneity on tracking accuracy are reported.
A tissue phantom demonstration is described, and we present the results of a usability study during which 12 clinicians with varying levels of experience used the device to assist with simulated focal liver lesion biopsies in an \textit{ex vivo} tissue phantom.

\begin{figure}[hbt]
	\centering
	\includegraphics[width=1.0\textwidth, trim=3.5cm 4cm 1cm 1cm, clip]{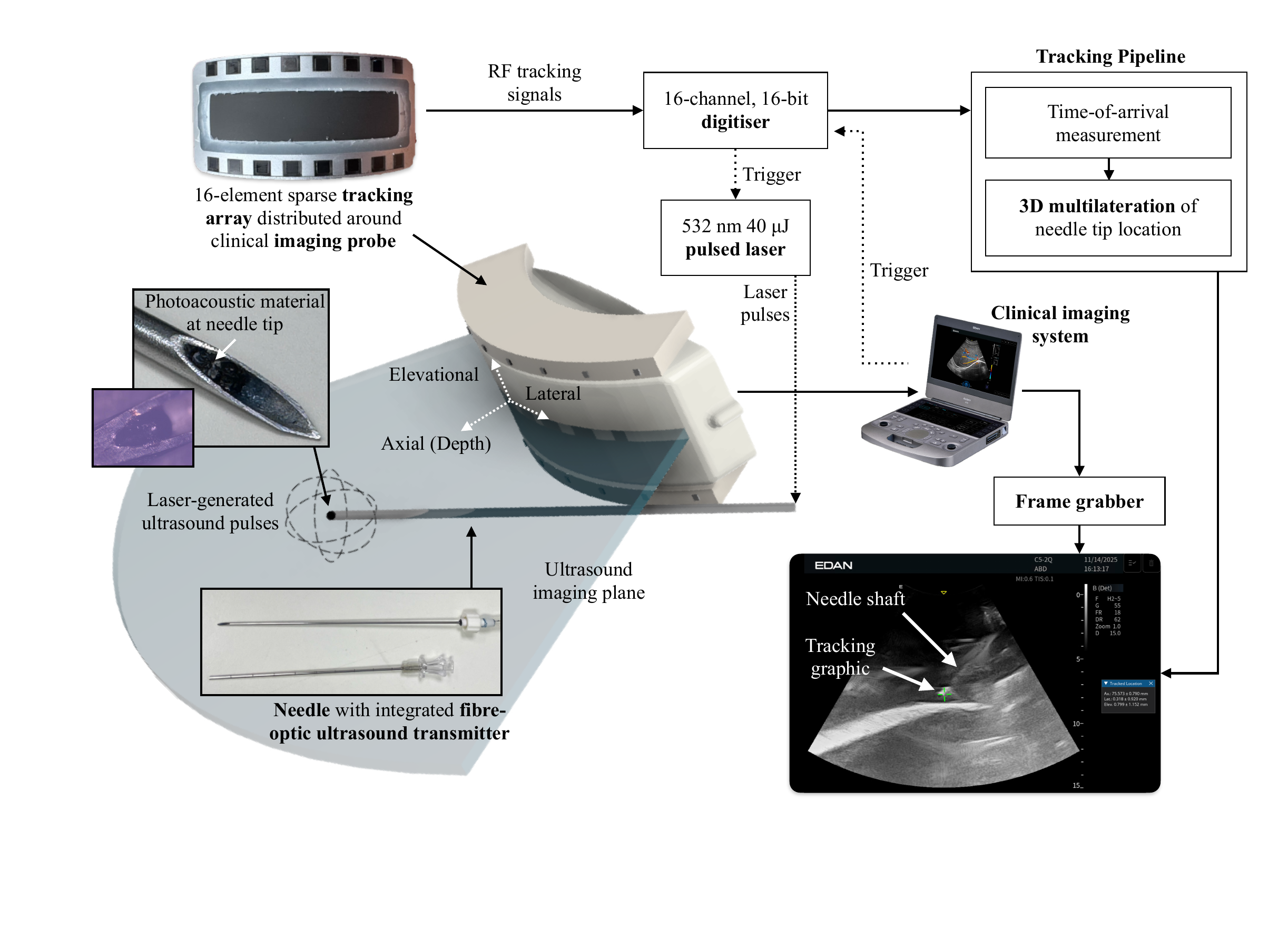}
	\caption{Conceptual graphic and schematic of the 3D ultrasonic needle tracking system, showing how pulses of ultrasound are generated from the needle tip and detected by a sparse receiver array, enabling 3D tracking of the needle tip within or either side of the imaging plane. Includes inset photographs of: the tip of the trackable needle with integrated fibre-optic ultrasound transmitter; the trackable needle stylet and blunt introducer needle; the tracking array attached to the ultrasound imaging probe; and the clinical ultrasound imaging system. }
	\label{fig:concept}
\end{figure}

\label{sec:method}
\section{Tracking and Imaging System}

The system achieves real-time 3D tracking by transmitting pulses of ultrasound from a fibre-optic ultrasound transmitter embedded within the needle tip (the "trackable needle").
The transmitter's photoacoustic coating is excited with a pulsed laser, generating pulses of ultrasound which are detected by an array of 16 piezoelectric receiving elements (the "tracking array") distributed around a 2--5\,MHz curvilinear clinical ultrasound probe.
This imaging probe is connected to a clinical ultrasound imaging system, from which imaging frames are streamed in real-time.
These imaging frames are annotated with a 3D visualisation of the needle tip's location, derived from the ultrasound signals received from the trackable needle by the tracking array.
A tracking console controls the generation of ultrasound, the acquisition of signals from the tracking array, and the streaming, annotation and display of the ultrasound imaging feed from the imaging system.
The imaging system provides a trigger signal to the tracking console synchronised with the end of its imaging frames, so that tracking pulses and imaging frames can be interleaved.

\subsection{Trackable Needle and Biopsy Device}

The single-use trackable needle was designed for compatibility with 18-gauge, 15\,cm TEMNO Evolution Biopsy Device (Merit Medical Systems, Utah, USA), replacing its solid inner stylet.
As sold, the biopsy device comprised: a blunt, 17-gauge, 10\,cm long ``introducer" needle; a solid, four-sided bevel, 18-gauge, 10\,cm long needle stylet; and a 15\,cm long core needle biopsy tool.
This gauge and length of tool were chosen after conversations with clinicians indicated it was the most commonly used.
During clinical practice, the stylet is initially fixed within the introducer needle with a Luer-lock hub.
The clinician first inserts this coaxial needle pair towards the target lesion under hand-held ultrasound guidance, positioning the bevelled tip at the surface of the lesion.
The inner stylet is then withdrawn, and the introducer needle used to guide the biopsy tool to and through the lesion, protruding from the end of the introducer needle.
This inner stylet was replaced with our trackable needle.
The introducer needle is shown in \autoref{fig:concept}.

Efficient PA generation requires an absorbing material with a high Gr{\"u}neisen parameter $\Gamma=\frac{\beta c^2_s}{C_p}$ ($\beta$ is the thermal expansion coefficient, $c_s$ is sound speed and $C_p$ is specific heat capacity at constant pressure) and optical absorption coefficient $\mu_a$ \cite{noimarkPolydimethylsiloxaneCompositesOptical2018}.
The fibre-optic transmitters were manufactured from 200\,\textmu m core, 245\,\textmu m outer-diameter, silica optical fibre (FG200-LEP, Thorlabs, New Jersey, USA).
A fibre was first cut to length (typically 1\,m), and then one end was connectorised with an 245\,\textmu m bore SMA905 connector (Thorlabs, New Jersey, USA).
The other end was then stripped of its polyamide coating, cleaved and dip-coated in a suspension of unfunctionalised reduced graphene oxide (rGO) powder (Sigma-Aldrich, MA, USA) in a 500\,g/L solution of PDMS (MED-1000, Avantor, Pennsylvania, USA) in xylene.
rGO provided a high $\mu_a$, while suitable thermal properties for PA generation were provided by the PDMS.
rGO was added with a 4\% mass fraction relative to the neat PDMS.
This mixture was prepared by manually mixing the constituent materials with a micro-spatula and then sonicating with a 3\,mm sonicator probe for 30 seconds (Fisherbrand Model 505/705 Sonic Dismembrator, Fisher Scientific, Leicestershire, UK).
During dip coating, the connectorised end of the fibre was connected to a fibre-coupled LED light source (FS201 Fiber Inspection Scope, Thorlabs, New Jersey, USA).
A single droplet of the prepared mixture was then deposited on to a glass slide, and the cleaved end of the optical fibre was repeatedly dipped into the droplet by hand until no light was visibly emitted from the dipped end of the fibre.
This typically took between 10 and 50 dips, and resulted in an approximately spherical coating with a diameter of 0.5--1\,mm.
This size, in combination with the optical absorption (i.e. ratio of rGO to PDMS) was chosen to reduce the acoustic output within the ultrasound imaging band (2--5\,MHz) while maintaining output at low frequencies, reducing the likelihood of interference from the trackable needle affecting the ultrasound image.
The coated fibres were then left to cure for 24 hours with the coated tips pointing vertically downwards.
A transmitter is shown within the needle bevel in \autoref{fig:concept}.

Each trackable needle was assembled from: a Hamilton (Nevada, USA) 14.5\,cm long, 18-gauge hubless needle with a standard 12$^\circ$ bevel; a Masterflex Nylon Male Luer Hub with 3.175\,mm hose barb fitting (Avantor, Pennsylvania, USA); 4\,mm outer-diameter silicone tubing; and a fibre-optic ultrasound transmitter.
A short length of 3.3\,mm outer-diameter polyvinyl chloride tubing was used to position the needle shaft centrally within the hub, with 10\,cm of needle shaft (excluding the bevel) protruding from the hub.
The fibre-optic ultrasound transmitter was fixed within the needle with epoxy, with its transmitting tip positioned at the base of the needle bevel as shown in \autoref{fig:concept}.
The silicone tubing was used to cover the length of fibre between the Luer hub and SMA connector, with the tubing fitted securely over the hose barb and attached to the SMA connector with self-amalgamating tape.

\subsection{Imaging System}
The imaging system, shown in \autoref{fig:concept}, was an Edan Acclarix AX8 with C5-2Q probe (Edan Instruments, Shenzhen, China).
The system was configured with a 65$^\circ$ beam angle and 15\,cm imaging depth.
A bespoke trigger output was provided by the manufacturer, which was connected to the tracking console for synchronisation of imaging and tracking.
The system's software was also modified by the manufacturer to enable control over the time gap between each imaging frame, which was configured to provide a frame rate of 10.5\,Hz.

\subsection{Tracking Array}
The tracking array, manufactured to our specification by Eintik Technologies (Shanghai, China), was designed to fit around the C5-2Q probe with a row of 8 elements positioned either side of and aligned parallel to the imaging array, as shown in \autoref{fig:concept}.
This design of 1\,MHz, 1\,mm $\times$ 1\,mm square elements was chosen to provide sufficient receive sensitivity across the field of view of the ultrasound imaging system to sub-arrays of at least 3 elements, given the directivity of the elements.
The choice of 16 elements allowed the use of a high-performance 16 channel digitiser without requiring multiplexing.
The choice of 1\,MHz, as well as widening directivity, also reduced the likelihood of acoustic interference from the ultrasound imaging system affecting tracking performance.
The separation between the two 8-element curvilinear sub-arrays was 33\,mm; this distance contributed to tracking accuracy in the elevational direction, and was chosen as a compromise between this and ergonomic and coupling considerations.
Within each of the 8-element curvilinear rows of the array, the element spacing was 9.71$^\circ$.
The tracking array was semi-permanently attached to the imaging probe using 2-part silicone rubber, as shown in \autoref{fig:concept}.

\subsection{Tracking Console}
The core component of the tracking console was a PC workstation housing a low-latency PCIe frame grabber (ProCapture HDMI, Nanjing Magewell Electronics Co., Ltd, Jiangsu Province, China) for acquisition of frames from the ultrasound imaging system, and hosting a high-performance 16-channel, 16-bit digitiser  (Digitizer Netbox, Spectrum Instrumentation GmbH, Grosshansdorf, Germany) via an Ethernet connection for acquisition of tracking waveforms, triggering of ultrasonic transmissions from the trackable needle and synchronisation with the imaging system.
Connections between these devices are shown in \autoref{fig:concept}.
On receipt of a trigger signal from the imaging system, the digitiser triggered a sequence of pulses from a 532\,nm, 40\,\textmu J, fibre-coupled, 5\,ns pulsed laser (CNI MPL-H-532, Changchun, China), to which the trackable needle was connected.
The laser, in turn, provided its own trigger signal synchronised to the generated laser pulses, and therefore the generation of ultrasound from the trackable needle.
This signal was used to trigger the acquisition of voltage waveforms from all 16 channels of the digitiser simultaneously with a sample rate of 8\,MHz.
Forty five ultrasound pulses were transmitted between each imaging frame, with the resulting waveforms averaged and used to generate a single tracked position, providing a tracking rate of 10.5\,Hz.

\subsection{Tracking Software}
\label{sec:software}
\begin{figure}[h]
\centering
\includegraphics[width=1.0\textwidth, trim=1cm 1.2cm 1cm 1.2cm, clip]{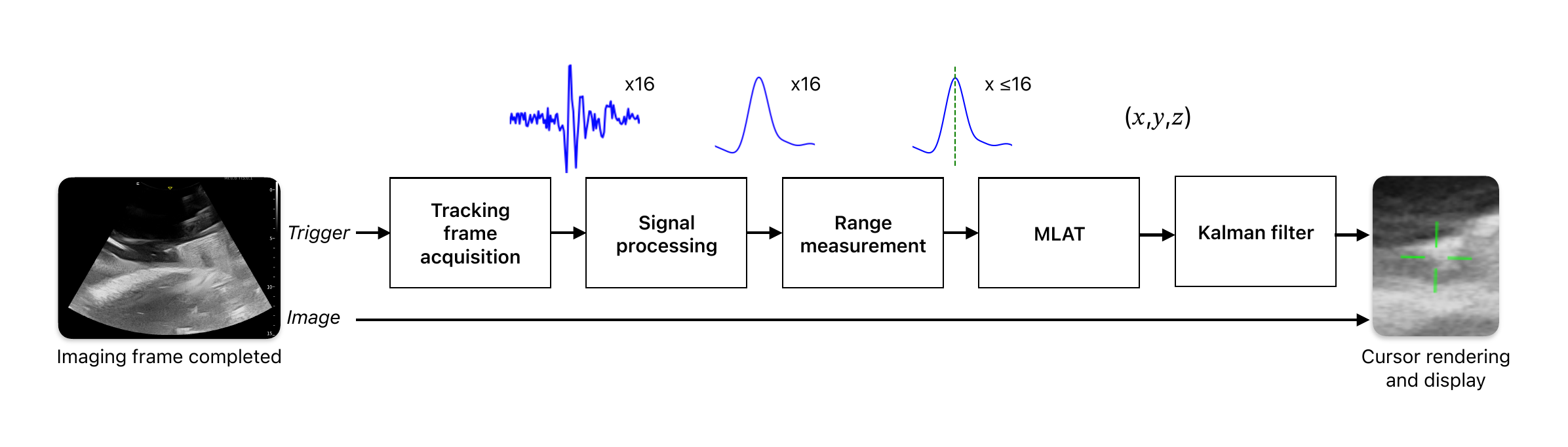}
\caption{Diagram of the tracking pipeline, showing how the acquisition of tracking waveforms is triggered by the completion of an imaging frame, and the waveforms are processed to determine the distance (range) between the needle tip and each element, and then ultimately a tracked position $(x,y,z)$, which is rendered on to a frame grabbed from the imaging system.}
\label{fig:pipeline}
\end{figure}

A multiprocessed, real-time, desktop application, written in Python 3.11, ran on the PC workstation. The software: multilaterated the tracked location of the needle tip $\bm{p}_{\text{tracked}}$ from the received waveforms; rendered $\bm{p}_{\text{tracked}}$ on top of the received ultrasound imaging frames; and displayed the rendered frame on the tracking console's monitor.
An overview of the tracking pipeline is shown in \autoref{fig:pipeline}.

Received, averaged waveforms were first high-pass filtered to remove any DC offset (200\,kHz cut-off).
A matched filter was then applied, with the template derived from measurements of typical high-SNR waveforms received by the tracking array from the trackable needle.
The output of the matched filter was then envelope-detected using the same method used for our previous 2D tracking system \cite{bakerIntraoperativeNeedleTip2022}.
Each waveform was then assessed for the presence of a pulse by measuring the skew of the waveform; waveforms with a skew less than 2 were discarded.

The time of arrival $t_i$ of the pulses in each remaining waveform was measured by finding the peak sample, and converted to a range $r_i$, assuming a speed of sound $c$ of either 1480\,m/s in water or 1540\,m/s in tissue:
\begin{equation}
r_i = c \cdot t_i.
\end{equation}
While sound speed heterogeneities encountered in tissue would certainly introduce tracking errors, it was hypothesised that ultrasound imaging, which also assumes homogenous sound speed and from which the patient anatomy is derived, would be similarly affected.
The purpose of the device was to track the location of the needle tip within the ultrasound image, rather than in physical space, and therefore a homogenous sound speed assumption was chosen.

Impossible ranges---those where the difference in the determined range between two neighbouring elements was larger than the physical distance between the elements---were ignored.
A second check ensured that the selected range measurements were distributed evenly across the two sides of the array: measurements from array elements that were not directly or diagonally opposite another element with its own valid range measurement were discarded.
The remaining ranges were stored; as were the amplitudes of the waveforms from which they were derived, for use during MLAT to estimate range measurement uncertainty.

MLAT was performed using Maximum Likelihood Estimation (MLE) to find the needle tip location $\bm{p}_{\text{tracked}}$ that maximised the likelihood of observing the measured ranges.
A cost function $J$ was minimised, equal to the negative logarithm of the posterior probability of an estimated needle location $\hat{\bm{p}}$, given the measured ranges $R=\{r_1,...,r_N\}$ and the last known location of the needle tip $\bm{p}_\textrm{prior}$~\cite{bishopPatternRecognitionMachine2006}.

For each location estimate $\hat{\bm{p}}$, a forward-model calculated the distance from $\hat{\bm{p}}$ to the known locations of each tracking element, $E = \{\bm{e}_1, ..., \bm{e}_N\}$:
\begin{equation}
r_{\hat{p},i} = ||\bm{\hat{p}}- \bm{e}_i||
\end{equation}
where $||\cdot||$ denotes the Euclidean norm.

The total likelihood $L$ of observing a set of ranges $R$ is the product of the probabilities of making each range measurement $r_i$, given the locations of the array elements (each being $\bm{e}_i$) and the estimated needle tip location $\bm{\hat{p}}$:
\begin{equation}
L(\bm{\hat{p}} | R, E) = \prod_{i=1}^{N} P(r_i | \bm{\hat{p}}, \bm{e}_i).
\end{equation}
The probability of a given range measurement occurring, $P(r_i | \bm{\hat{p}}, \bm{e}_i)$, was modelled as a mixture of two normal distributions, $\mathcal{N}(\mu, \sigma^2)$:
\begin{equation}
P(r_i | \bm{\hat{p}}, \bm{e}_i) = w_{1,i} \mathcal{N}(r_i | r_{\hat{p},i}, \sigma_{1,i}^2) + w_{2,i} \mathcal{N}(r_i | \mu_{2,i}, \sigma_{2,i}^2).
\end{equation}
Here, the first normal distribution represents the probability of obtaining the measurement $r_i$ if the received pulse was correctly detected (i.e. if the peak sample of the waveform lay within the pulse); it is therefore centred on the forward-modelled range $r_{\hat{p},i}$.
The second distribution represents the probability of the measured range occurring if the pulse was missed, for instance due to a low signal-to-noise ratio; it is centred on a value $\mu_{2,i}$ corresponding to the centre of the waveform acquisition window.
An example of a resulting probability density function is shown in the Supplementary Materials.
The standard deviations $\sigma_{1,i}, \sigma_{2,i}$ and weights $w_{1,i}, w_{2,i}$ for each measurement's probability model were generated dynamically from the amplitude of the corresponding received waveform; functions relating received signal amplitude to these parameters were determined from empirical assessment of the waveforms received from known positions (and therefore with known true ranges). See Supplementary Materials for details.

A Gaussian prior on the needle tip location was included, derived from its last tracked location $\bm{p}_\textrm{prior}$, the tracking rate and expected needle velocity.
The corresponding penalty in the cost function was the negative log-likelihood of this prior, a standard approach for incorporating prior knowledge in Bayesian estimation~\cite{bishopPatternRecognitionMachine2006}.
It was assumed that the uncertainties on the prior location in each axis were independent and had equal standard deviation $\sigma_{\text{prior}}$, which was set to 6\,mm, effectively assuming that the needle tip was unlikely to move more than this distance between each tracking frame.

The cost function $J$ was therefore defined as:
\begin{equation}
J(\bm{\hat{p}}) = - \log L(\bm{\hat{p}} | R, E) + \frac{1}{2\sigma_{\text{prior}}^2} ||\bm{\hat{p}} - \bm{p}_{\text{prior}}||^2.
\end{equation}
The final tracked location was the estimate $\hat{\bm{p}}$ that minimised this cost function:
\begin{equation}
\bm{p}_{\text{tracked}} = \argmin_{\hat{\bm{p}}} J(\hat{\bm{p}}).
\end{equation}
Minimisation was performed using the Scipy.Optimize Python package \cite{virtanenSciPy10Fundamental2020} using the L-BFGS-B algorithm \cite{byrdLimitedMemoryAlgorithm1995}. 
The cost function was implemented using the JAX Python library \cite{47008} enabling just-in-time compilation for real-time tracking and automatic determination of the Jacobian, which aided minimisation.

A Kalman filter was implemented to optimally combine new tracked positions with a prediction of the location of the needle tip based on its previous location and velocity.
The state of the needle tip at each time step was represented by a vector containing its position and velocity.
The filter predicted the next state using a constant-velocity model.
The process noise covariance was constructed assuming a piecewise constant acceleration over each time step with a process noise parameter of 0.01\,m$^2$/s$^4$.
This parameter represented the variance of acceleration magnitude, indicating to the filter the uncertainty in the predictions of the constant velocity model.

\begin{figure}[h]
	\centering
	\begin{subfigure}[t]{0.57\textwidth}
		\vspace{0pt} 
		\centering
		\includegraphics[width=\textwidth, trim=7cm 1cm 7cm 2cm, clip]{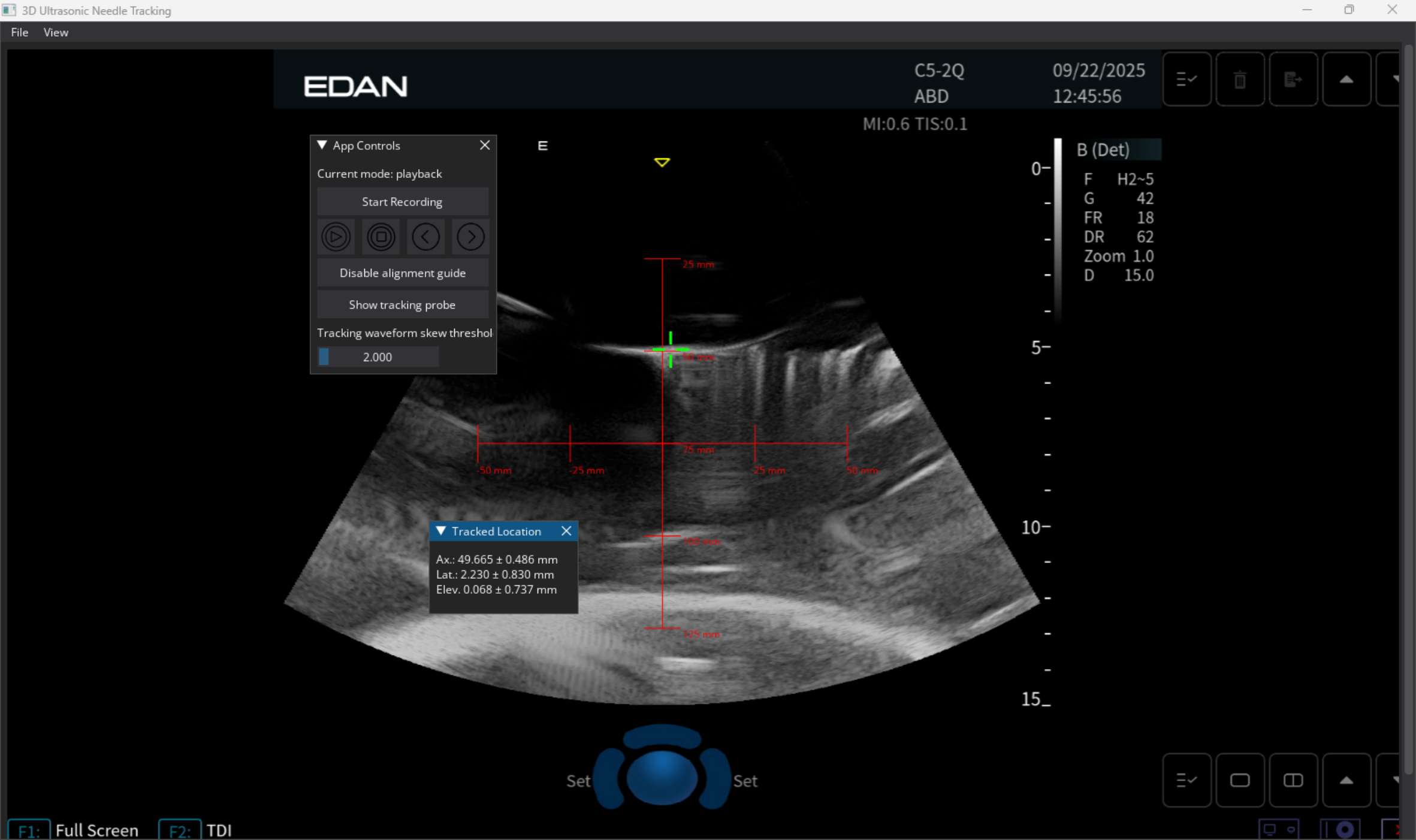}
		\subcaption{Registration cross-hair}
		\label{fig:reg_crosshair}
	\end{subfigure}
	\hfill
	\begin{minipage}[t]{0.42\textwidth}
		\vspace{0pt} 
		\centering
		\begin{subfigure}[t]{\textwidth}
			\centering
			\includegraphics[trim=2cm 23cm 16.65cm 3.25cm, clip, width=\textwidth]{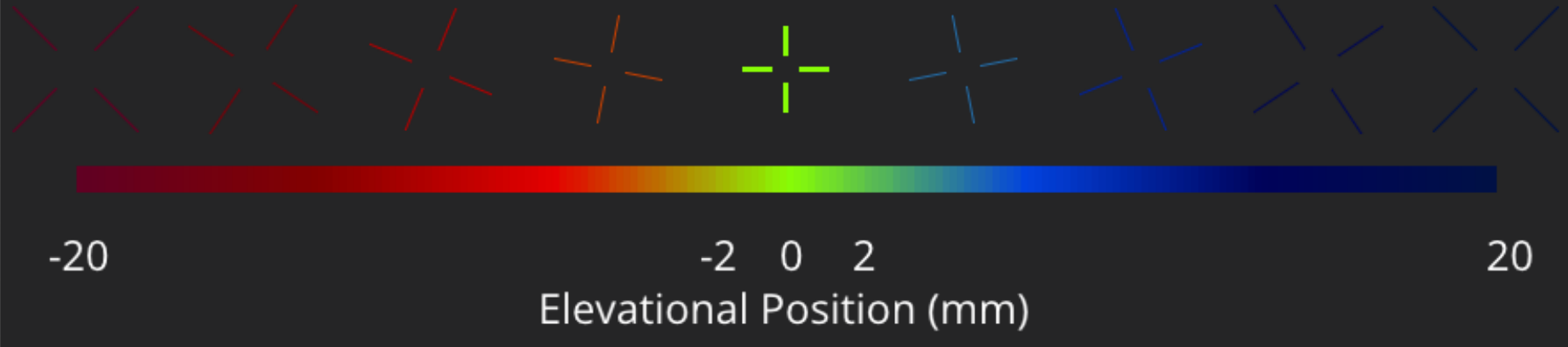}
			\subcaption{Usability study 1 cursor map}
			\label{fig:cursor_map1}
		\end{subfigure}\\[20pt]
		\begin{subfigure}[t]{\textwidth}
			\centering
			\includegraphics[trim=2cm 23cm 16.65cm 3.25cm, clip, width=\textwidth]{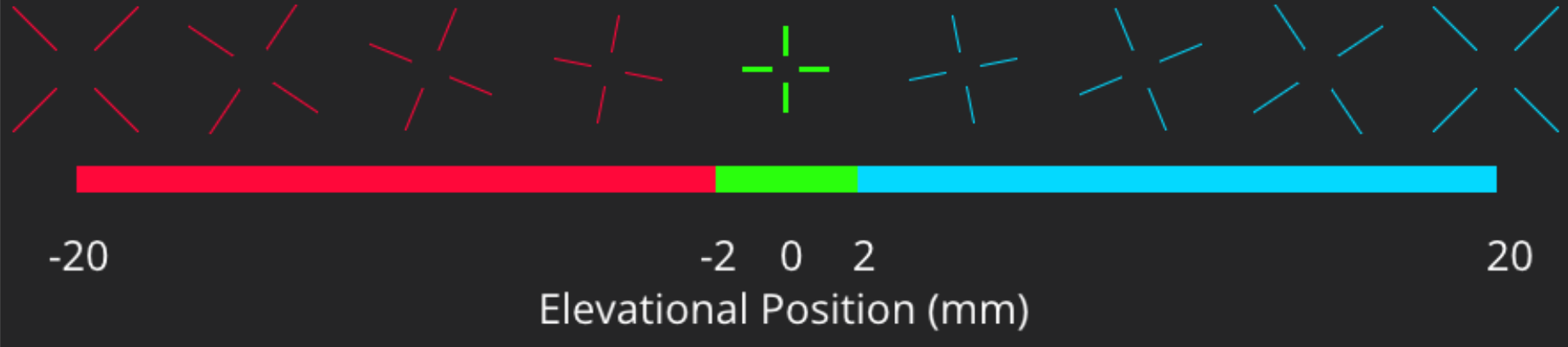}
			\subcaption{Usability study 2 cursor map}
			\label{fig:cursor_map2}
		\end{subfigure}
	\end{minipage}
	
	\caption{Visualisations related to registration and cursor rendering:
		(a) screenshot of the tracking software taken during registration, showing the unregistered tracked needle tip location (green) and the registration cross-hair (red); 
		(b) the tracking cursor colour and orientation map used for the first usability study; 
		(c) the tracking cursor colour and orientation map used for the second usability study.}
	\label{fig:combined_cursor_figures}
\end{figure}

Each 3D tracked location, $ \bm{p}_{\text{tracked}} = (x, y, z) $, was rendered on to the ultrasound imaging video feed by first converting the lateral $x$ and axial $y$ coordinates to a pixel location, $(u, v)$, within the ultrasound image.
This conversion was completed using knowledge of the location of the centre of the probe face within the image, $(u_c, v_c)$, and the size of the square pixels in millimetres, $s$:
\begin{equation}
	u = u_c + \frac{x}{s}, \quad v = v_c + \frac{y}{s}.
\end{equation}
A tracking cursor was then drawn, centred on this pixel, with its colour, size and orientation set proportional to the elevational distance of the needle tip from the imaging plane, $z$, as shown in \autoref{fig:combined_cursor_figures}.

\subsection{Registration}
\label{sec:registration}
Determination of the tracked location of the needle $ \bm{p}_{\text{tracked}} $ relative to the centre of the imaging probe, in coordinates aligned to the axial, lateral and elevational axes of the imaging system, required knowledge of the location of the tracking elements $E$ in that same coordinate system.
While the tracking probe was attached to the imaging probe with the intention of aligning its face flush with that of the imaging probe---which if performed perfectly, would negate the need for registration---in practice, it was necessary to account for any rotation, $\mathbf{R}$, and translation, $\bm{t}$, between the coordinate systems of the two devices.
This was performed acoustically by comparing a set of tracked locations of the needle tip---generated assuming the two coordinate systems were perfectly aligned---with the apparent location of the needle within the ultrasound image, at a set of $N$ predefined locations $P_\text{reg} = \{\bm{p}_{\text{reg},1}, ..., \bm{p}_{\text{reg},N}\}$ within the imaging plane, in water.
This method was chosen as it provided direct registration with ultrasound imaging, rather than relying on an additional tracking tool such as an optical or electromagnetic system which would itself require registration and validation.

For the purpose of registration, a 100\,mm $\times$ 100\,mm cross-hair was displayed on top of the ultrasound imaging feed at the centre of the field of view (see \autoref{fig:combined_cursor_figures}).
Using a 3-axis motion control system (1N150 linear stages, Thorlabs, New Jersey, United States), with the ultrasound probe assembly mounted with its face immersed in a water tank pointing vertically downwards, the needle tip was positioned at each of the 9 vertices marked on the cross-hair, visually aligning the tip to the vertices by examination of the ultrasound image, ensuring that the needle was in-plane by observing the brightness of the needle in the image.
At each location, a sequence of tracked locations (typically 30) were recorded.

For each location, the average of the tracked positions, $ \bm{\bar{p}}_{\text{tracked}, i} $ was compared to the known position of the vertex $ \bm{p}_{\text{reg}, i} $ in the imaging coordinate system, which was first corrected to account for the difference in sound-speed between the registration medium (water) and that assumed by the imaging system.
The assumed orientation, $\mathbf{R}$, and location, $\bm{t}$, of the tracking array relative to the imaging probe, which determined the tracking element locations used by the tracking algorithm, were then adjusted manually to minimise the mean-square error (MSE) between the known and tracked locations:
\begin{equation}
\text{MSE} = \frac{1}{N} \sum_{i=1}^{N} || \bm{p}_{\text{reg}, i} -  \bm{\bar{p}}_{\text{tracked}, i}(\mathbf{R}, \bm{t}) ||^2.
\end{equation}
The final orientation and translation was then stored and used for all future tracking measurements.
It was only required to complete this process once: after the semi-permanent assembly of the tracking array and imaging probe.

\section{Experimental Methods}

\subsection{Tracking Accuracy in Water}
\label{sec:accuracy_in_water_method}
Tracking accuracy was assessed in water using the same 3-axis motion control system and water tank used for registration.
The experiment, which included the steps described below, was repeated twice using the same trackable needle, on different days.
The resulting accuracy vectors were averaged across the two scans and their standard deviation taken to assess experimental repeatability.
The needle was held horizontally with its bevel facing upwards, completely immersed in room-temperature water.
The tracking and imaging probe assembly was fixed at the surface of the water, with its face immersed and pointing vertically downwards.
An ultrasonic absorber tile (F28, Precision Acoustics Limited, UK) was placed at the base of the water tank directly below the imaging probe, angled to further reduce the effect of reflections.

Ground truth positions were derived by determining the vectors---in the coordinate system of the motion stages---defining the axial, lateral and elevational axes of the imaging probe, and then moving the needle tip in a rectilinear grid aligned to these vectors.
This was achieved by moving the needle tip (under computer control) to 9 known in-plane locations within the ultrasound field of view using the same cross-hair overlay used for registration, shown in \autoref{fig:combined_cursor_figures}, ensuring the needle tip was precisely in-plane at each vertex by observing the brightness of the needle image.
Similarly to registration, this method provided direct alignment to the images that the tracking system was designed to annotate.
 In the reconstructed ultrasound imaging space, the centre of the cross-hair was 75\,mm directly below the imaging probe; assumed a speed of sound of 1540\,m/s rather than the true speed of sound of 1480\,m/s, and therefore the centre of the cross hair corresponded to a location 72.1\,mm physically below the imaging probe.
With the needle tip at each of the 9 locations, the coordinates of the three stages in the motion-controller coordinate space were stored.
Principal component analysis (PCA) was then used to determine initial vectors aligned to the horizontal ($\bm{v}_h$) and vertical ($\bm{v}_v'$) axes of the cross-hair in motion controller space.
The Gram-Schmidt process was used to adjust the vertical axis vector to be perfectly orthogonal to the horizontal axis vector, resulting in orthogonal vector $\bm{v}_v$.
The third, orthogonal (elevational) vector $\bm{v}_e$ was then calculated as the cross product of these two orthogonal vectors, i.e., $\bm{v}_e = \bm{v}_h \times \bm{v}_v$.
This information enabled the conversion of ultrasound imaging space coordinates to motion-controller coordinates.

Tracked positions were recorded at 342 locations in three axial-lateral planes 0, 10, and 20\,mm from the imaging plane, respectively.
Within each plane, locations were spaced laterally and axially by 10\,mm.
The three planes had lateral and axial extent completely covering the right hand side of the field of view of the ultrasound imaging system, and extending 10\,mm into the left hand side.
The maximum depth was 140\,mm.
Locations outside of the field of view of the imaging system were excluded.
At each location, $M=40$ tracked positions $ \bm{p}_{\text{tracked}} $ were recorded, which corresponded to 4\,s of acquisition time at the 10\,Hz tracking rate.
The mean, $\bm{\mu}_{\text{tracked}}$, and standard deviation, $\sigma_{\text{tracked}}$, of the tracked positions were calculated to assess tracking error and repeatability respectively:
\begin{equation}
 \bm{\mu}_{\text{tracked}} = \frac{1}{M} \sum_{j=1}^{M}  \bm{p}_{\text{tracked}, j},
\end{equation}

\begin{equation}
\sigma_{\text{tracked}} = \sqrt{\frac{1}{M-1} \sum_{j=1}^{M} || \bm{p}_{\text{tracked}, j} -  \bm{\bar{p}}_{\text{tracked}} ||^2}.
\end{equation}

\subsection{Tracking Accuracy in an \textit{ex vivo} Tissue Phantom}
Accuracy in tissue was assessed by placing a trackable needle within a simple \textit{ex vivo} tissue phantom, acquiring tracking data and then imaging the setup (i.e. phantom, inserted needle and mounted imaging probe) using CT to determine the true location of the needle tip relative to the imaging probe.

To allow measurement of accuracy at depths of 10\,cm and above, a one-off 15\,cm long trackable stylet was manufactured; due to the angle of approach, the 10\,cm trackable stylets could typically only reach depths of between 8 and 9\,cm.

The phantom comprised a 22\,cm $\times$ 11\,cm $\times$ 16\,cm plastic container filled with chicken breast and deionised water.
The chicken breast and water were added to the container layer by layer to prevent air becoming trapped between the tissue layers.

The phantom was placed within the field of view of a Medtronic O-arm CT gantry (Minnesota, USA) with the tracking and imaging probe assembly fixed with its face coupled to the surface of the phantom, pointing vertically downwards.
The needle was inserted to position its tip at 11 arbitrary locations within the volume of interest, attempting to keep the tip within 20\,mm of the imaging plane.
At each location, 100 tracked positions were recorded, immediately before taking a volumetric CT scan of the whole setup.
The resulting CT data were exported as DICOM files for analysis.
A photograph of the phantom within the CT gantry is provided in the Supplementary Materials.

The DICOM files were analysed in Python using the pydicom, pyvista and simpleITK packages. The phantom and ultrasound probe did not move during the data acquisition period, either relative to each other or relative to the CT gantry; this allowed the location of the imaging probe to be determined from only one dataset.
The volume was first manually cropped to remove the tracking array from the image so that only the region directly around and including the imaging probe face remained.
This cropped volume was then binarised using a threshold of 2000 Hounsfield Units (HU) to discard low-density voxels representing chicken tissue and water, leaving only those representing the imaging probe, and then a surface mesh was extracted.
A cylindrical section of the known shape and size of the imaging probe face was manually fitted to the mesh, and the final location ($\bm{p}_{\text{probe}}$) and orientation of the cylindrical section stored.
The orientation was defined by three orthogonal vectors $\bm{v}_{\text{lat}}$, $\bm{v}_{\text{ele}}$ and $\bm{v}_{\text{ax}}$, aligned with the probe's lateral, elevational and axial axes, respectively.
An image of the segmented volumes is provided in the Supplementary Materials.

The CT dataset acquired for each needle position was automatically processed to determine the location of the tip within the image.
Each volume was first cropped to remove the tracking and imaging probe assembly so that only voxels representing the needle and phantom remained.
It was then thresholded using a threshold of 2000 HU to select voxels likely to represent the needle.
PCA was then used to find the connected cluster of needle-like voxels with the largest variance ratio for its first principal component, $\bm{v}_1$, which indicated that the voxels of the cluster lay along a straight line and therefore were highly likely to represent the needle.
The coordinates of these voxels, each represented by $\bm{p}_i$, were then processed to find the tip: first, the principal component vector was oriented to point downwards, in the direction of needle insertion, to give $\bm{\hat{v}}_1$; then the voxel coordinates were centred by subtracting their centroid $\bm{\bar{p}}$; and finally the centred coordinates were projected onto the oriented vector $\bm{\hat{v}}_1$, generating a set of scalar distances $ S = \{s_1, ..., s_N\} $:
\begin{equation}
s_i = (\bm{p}_i - \bm{\bar{p}}) \cdot \bm{\hat{v}}_1.
\end{equation}
The location of the needle tip within the image $ \bm{p}_{\text{image}} $ was then determined from the maximum of these distances $s_k = \max(S)$.
To account for the position of the ultrasound transmitter within the needle, $ s_k $ was first adjusted by half the known length of the needle bevel $d_{\textrm{bevel}}=6$\,mm.
Finally, the location of the needle tip was calculated by multiplying the adjusted $s_k$ by the oriented principal component vector $ \bm{\hat{v}}_1 $, and then adding the centroid $ \bm{\bar{p}} $:
\begin{equation}
\bm{p}_{\text{image}} = (s_k - \frac{d_{\textrm{bevel}}}{2}) \bm{\hat{v}}_1 + \bm{\bar{p}}.
\end{equation}

For each needle placement, the displacement vector between the CT-derived needle tip coordinate, $ \bm{p}_{\text{image}} $, and the centre of the probe face, $ \bm{p}_{\text{probe}} $, was calculated in voxels.
This was then scaled by the known pixel size of the CT image, $ \bm{r}_{\text{CT}} $, to obtain the displacement in millimetres.
The dot product of this vector with each of the three vectors representing the orientation of the imaging probe ($ \bm{v}_{\text{ax}} $, $ \bm{v}_{\text{lat}} $, $ \bm{v}_{\text{ele}} $) was then calculated to determine the location of the needle tip relative to the imaging probe $ \bm{p}_{\text{CT}} $, within the imaging probe's own coordinate system.
The tracked $ \bm{p}_{\text{tracked}} $ and CT-derived $ \bm{p}_{\text{CT}} $ locations were then compared by calculating the error vector, $ \bm{e} $:
\begin{equation}
\bm{e} = \bm{p}_{\text{tracked}}-\bm{p}_{\text{CT}}.
\end{equation}
The standard-deviation of the tracked positions recorded at each location was also calculated to assess tracking repeatability.

\subsection{Simulation Study of the Effects of Sound Speed Heterogeneity}

\begin{figure}
	\begin{subfigure}[b]{0.48\textwidth}
		\centering
		\includegraphics[width=\textwidth]{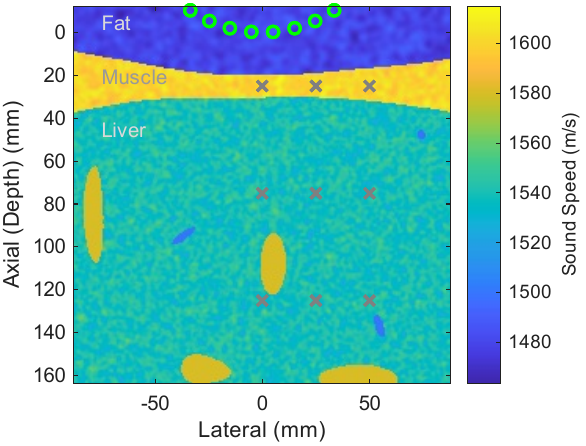}
		\subcaption{}
	\end{subfigure}
	\begin{subfigure}[b]{0.48\textwidth}
		\centering
		\includegraphics[trim=3cm 9cm 5cm 8.5cm, clip, width=0.9\textwidth]{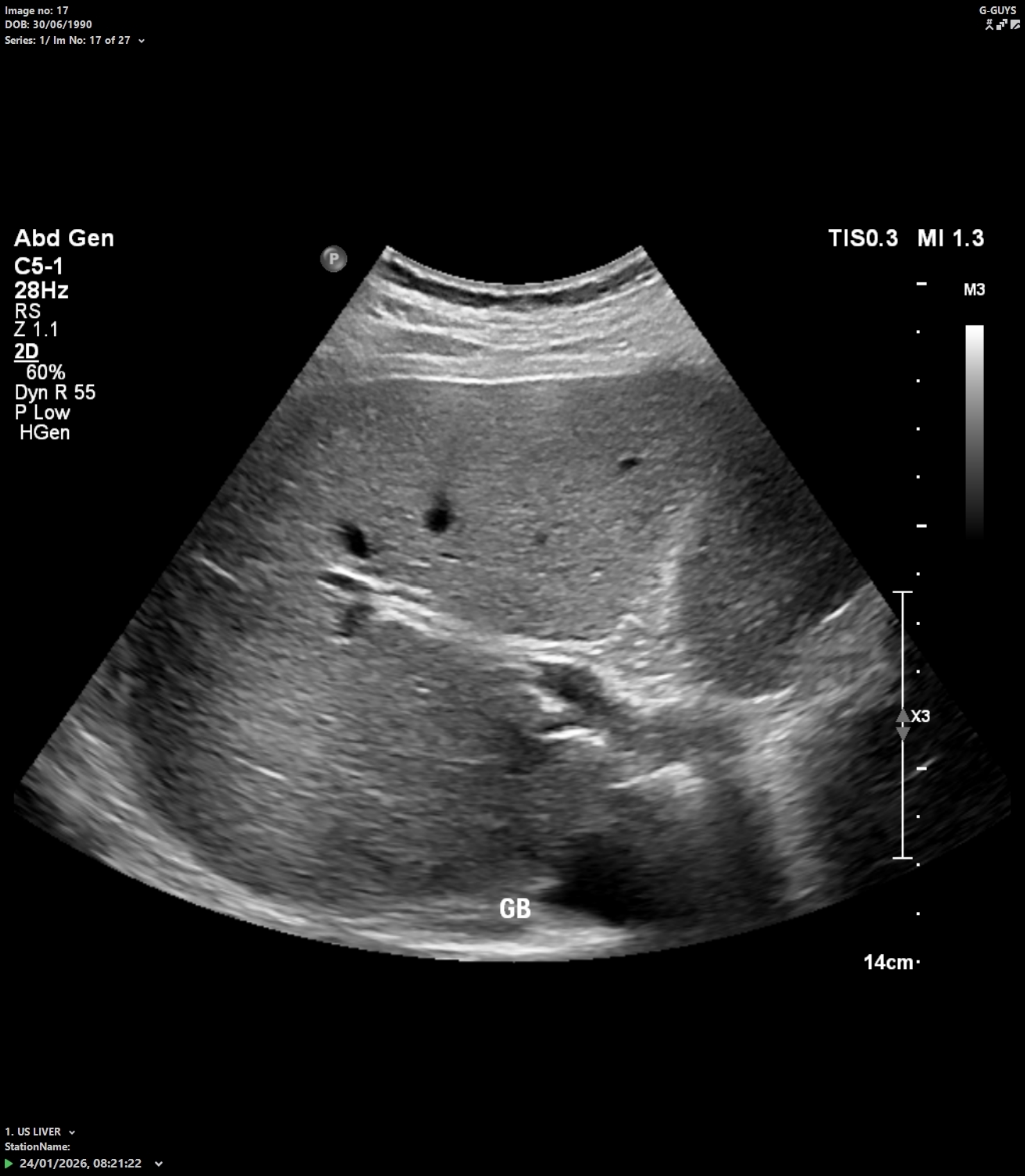}
		\subcaption{}
	\end{subfigure}
	\caption{(a) Central cross-section of one of the ten randomly-generated digital sound-speed phantom used for ultrasound simulations, showing the locations of the tracking elements (green circles) and trackable needles (grey crosses), and annotated to show the three tissue layers. For comparison, an image from a clinical liver ultrasound is shown in (b).}
	\label{fig:simulation_setup}
\end{figure}

The k-Wave \cite{treebyKWaveMATLABToolbox2010} MATLAB (The MathWorks, MA, USA) Toolbox was used to simulate the effects of sound speed heterogeneity on tracking.
Ten 178\,mm $\times$ 45\,mm $\times$ 178\,mm 3D digital sound speed phantoms were procedurally generated in MATLAB to mimic the tissues beneath the ultrasound probe during ultrasound examination of the liver.
Each phantom comprised three layers representing subcutaneous fat, the rectus abdominus and the liver.
Layers were randomly curved to mimic the effect of the pressure of the curvilinear ultrasound probe in the skin and the natural curvature of the liver.
The liver layer included randomised tubular structures representing three blood vessels and two bile ducts.
Details of the phantom construction can be found in the supplementary materials.
A central slice of one of the ten phantoms is shown in \autoref{fig:simulation_setup} next to an image taken from a clinical liver ultrasound, for comparison.

Layer thicknesses and vessel sizes were chosen based on published data \cite{choiRelationshipRectusAbdominis2017, kimThicknessRectusAbdominis2012, storchleMeasurementMeanSubcutaneous2018, draghiUltrasoundExaminationLiver2007} and examination of available abdominal ultrasound images.
The nominal layer thicknesses and vessel sizes used for each of the ten phantoms were randomly chosen from normal distributions with means and standard deviations shown in \autoref{tab:digital_phantom}.
For the purposes of simulation, the tracking array was located within the fat layer, requiring an additional 15\,mm of nominal thickness.

The average sound speed in each layer is shown in \autoref{tab:digital_phantom}.
Heterogeneity within the tissue layers was generated using Perlin noise.
Sound speeds were chosen from published data \cite{duckPhysicalPropertiesTissues1990, wangUltrasonicSoundSpeed2023a}.
Heterogeneity was randomised for each of the ten phantoms.

Each digital phantom was used to simulate the propagation of ultrasound from 18 simulated needle locations and the known locations of the tracking array elements, which were situated at the top of the phantom within the fat layer.
The needle locations were in two axial-lateral planes 0 and 10\,mm from the imaging plane, respectively.
Each plane contained a 3\texttimes3 grid of simulated needle locations, with each location separated by 25\,mm laterally and 50\,mm in depth.

The needle was simulated as a point source excited with an impulse, and the time domain acoustic pressure waveform received at each array element location was recorded, generating a simulated tracking frame.
Details of the simulation can be found in the supplementary materials.
These tracking frames were saved to disk and then processed in Python using the tracking pipeline described in \autoref{sec:software} to generate 10 tracked locations per position (one for each digital phantom).
The tracking pipeline included a matched filter representing the system response, and as such it was not necessary to simulate the acoustic response of the fibre-optic ultrasound transmitter or tracking array elements.
Tracking error and repeatability were assessed as described in \autoref{sec:accuracy_in_water_method}.

\begin{table}[h]
	\centering
	\caption{Properties of the digital sound speed phantom used to assess the effects of heterogeneity on tracking accuracy. Nominal thickness indicates the layer thickness prior to the addition of curvature and noise. Nominal thicknesses were randomised for each phantom from normal distributions, and the values in this table are the means and standard deviations of these distributions. Sound speed was randomised within each tissue layer of each phantom; the means and variances of the Perlin noise used for this randomisation are shown here. Vessel and duct sound speeds were homogenous and constant across all phantoms, but vessel diameter ("thickness") was randomised between phantoms using normal distributions defined by the means and standard deviations shown here.}
	\label{tab:digital_phantom}
	\begin{tabular}{l c c}
		\toprule
		& \textbf{Nominal Thickness (mm)} & \textbf{Sound speed (m/s)} \\
		\midrule
		Fat & $25 \pm 5$ & $1480 \pm 20$ \\
		Muscle & $10 \pm 5$ & $1600 \pm 20$ \\
		Liver & - & $1540 \pm 30$ \\
		Blood vessel & $12 \pm 3$ & $1580$ \\
		Bile duct & $6 \pm 2$ & $1500$ \\
		\bottomrule
	\end{tabular}
\end{table}

\subsection{\textup{Ex vivo} Tissue Phantom Demonstration}
To demonstrate the real-time tracking capability, an insertion was carried out by one of the authors into an \textit{ex vivo} tissue phantom.
The phantom was constructed as for the tracking accuracy experiment described above, and then a 2.5\,cm $\times$ 2.5\,cm $\times$ 2.5\,cm cube of beetroot was placed between the chicken breast layers at a depth of approximately 5\,cm to simulate a target lesion.
The needle was inserted into the phantom with the aim of reaching the needle tip to the surface of the lesion.
A screen recording of the tracking software interface was taken during the insertion.
The resulting video was annotated to show the location of the lesion and needle shaft (when visible) and captioned to describe the motion of the ultrasound probe assembly and needle throughout.

 \subsection{Usability Tests in an \textup{ex vivo} Tissue Phantom}
Twelve interventional radiologists completed simulated lesion biopsies on \textit{ex vivo} tissue phantoms during two studies carried out at St Thomas' Hospital (London, UK).
Biopsies were performed both with and without tracking.
The time taken to complete each biopsy was recorded, and its success was determined by visual examination of the extracted material.
Between the studies, the cursor colour map was changed from a continuous map that ranged between a dark red and a dark blue, to the three-colour map described in \autoref{sec:software}, based on feedback received during the first study.
The two colour maps are shown in \autoref{fig:combined_cursor_figures}.

Five clinicians attended the first usability study: two consultant interventional radiologists, and three trainees in their fifth-year, second-year and first-year respectively.
A further seven clinicians attended the second study: three consultants, a clinical fellow (specialist registrar), two sixth-year trainees and a fourth-year trainee.

\begin{figure}[h!]
	\centering
	
	\begin{subfigure}[b]{0.35\textwidth}
		\centering
		\begin{minipage}[c][7cm][c]{\textwidth}
			\centering
			\includegraphics[width=\textwidth,
			trim=1.4cm 1.5cm 1.3cm 0cm, clip]{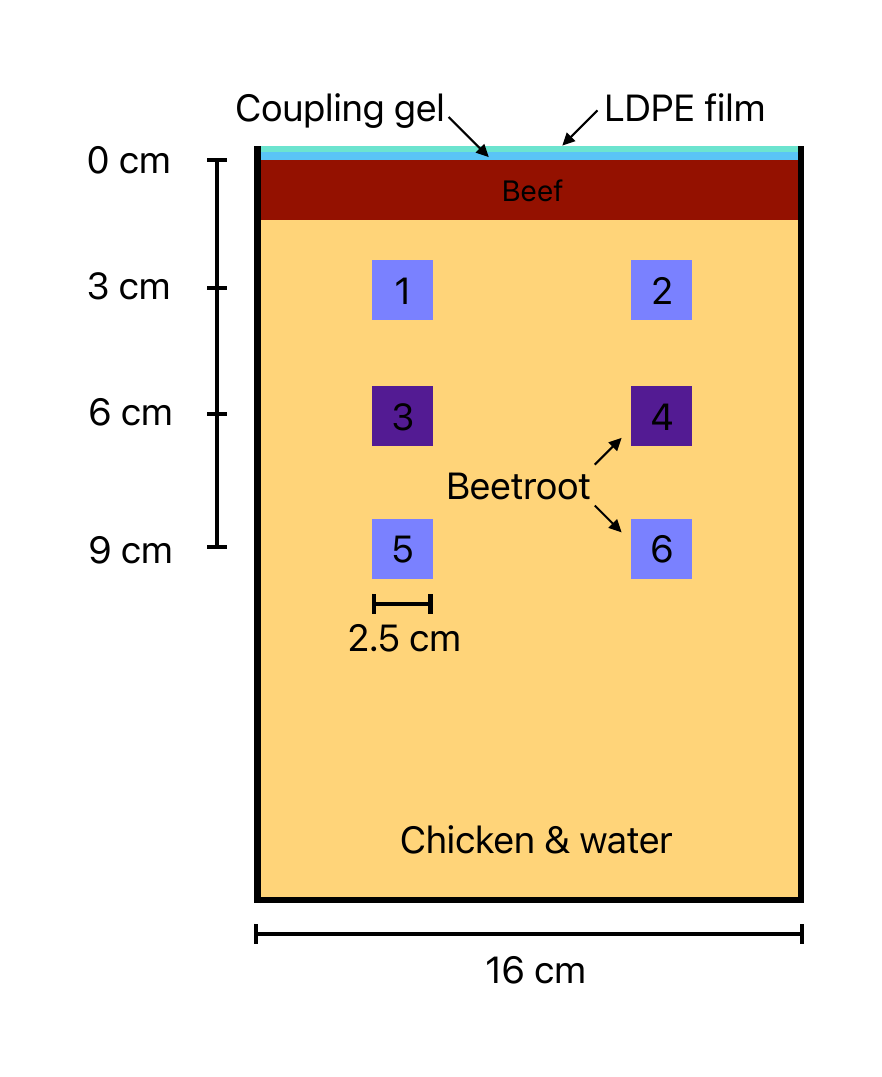}
		\end{minipage}
		\subcaption{}
		\label{fig:phantom_diagram_a}
	\end{subfigure}%
	\hfill
	\begin{subfigure}[b]{0.29\textwidth}
		\centering
		\begin{minipage}[c][6.6cm][c]{\textwidth}
			\centering
			
			\begin{tikzpicture}
				\node[anchor=south west, inner sep=0] (img) 
				{\includegraphics[width=\textwidth, viewport=129 120 540 630, clip=true]{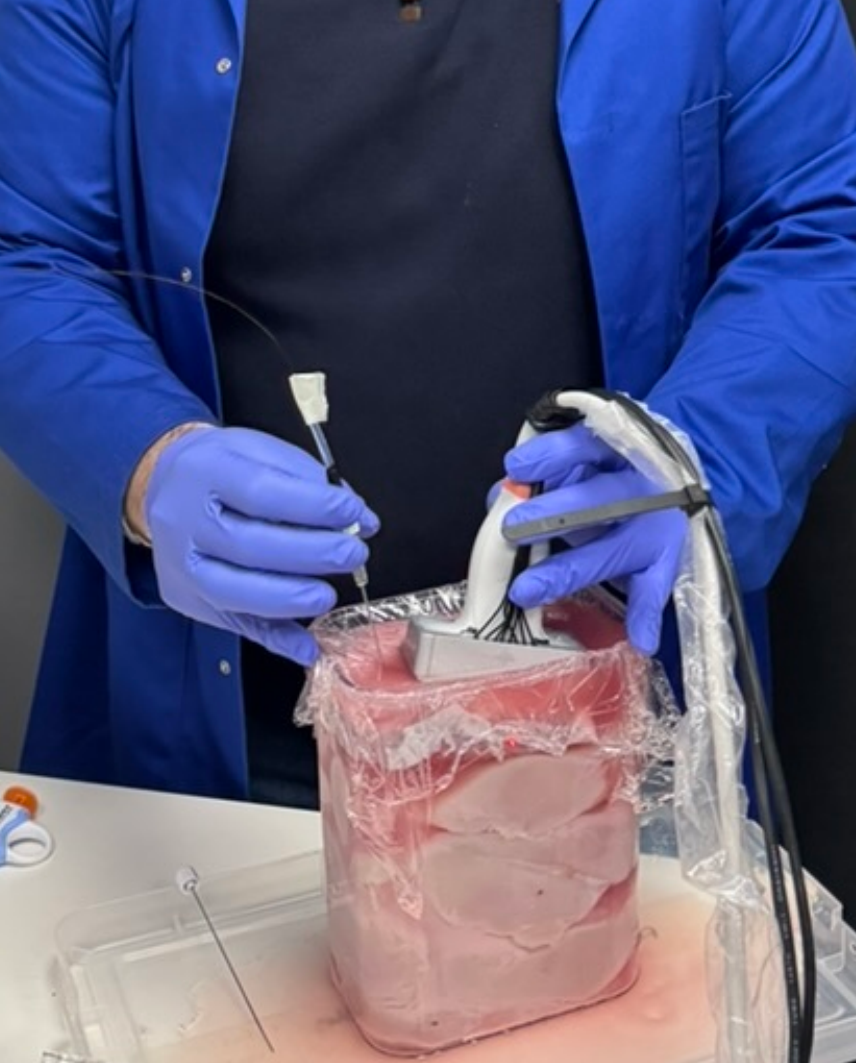}};
				
				\node[white, font=\bfseries] at (2.6,0.8) {Phantom};

				\draw[white, thick, ->] 
				(2.5, 3.85) -- (2.6,3.2);
				\node[white, font=\bfseries, align=center] at (2.5, 4.3) {Probe\\assembly};
				
				\draw[white, thick, ->] 
				(0.9, 2.2) -- (1.7, 3.2);
				\node[white, font=\bfseries] at (0.9, 2) {Needle};
				
			\end{tikzpicture}
			
		\end{minipage}
		\subcaption{}
		\label{fig:phantom_diagram_b}
	\end{subfigure}%
	\hfill
\begin{subfigure}[b]{0.32\textwidth}
	\centering
	\begin{minipage}[c][6.625cm][c]{\textwidth}
		\centering
		
		\begin{tikzpicture}
			\node[anchor=south west, inner sep=0] (img) 
			{\includegraphics[width=\textwidth,
				trim=15cm 17cm 27cm 3.4cm, clip]{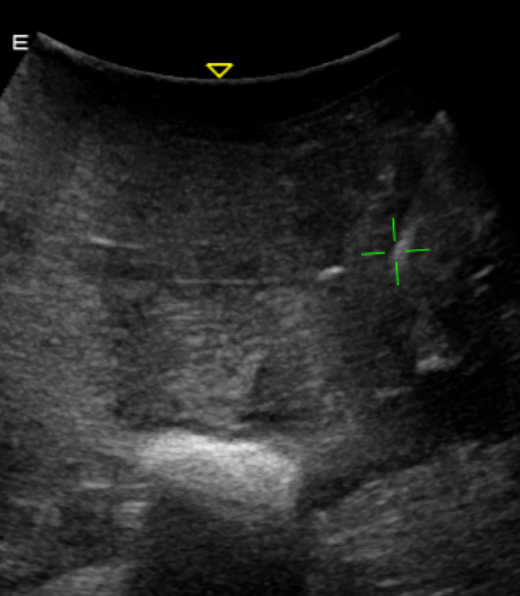}};
			
			\draw[white, thick, ->] 
			(2.5,3.8) -- (3.5,3.4);
			\node[white, font=\bfseries] at (2.5,4.0) {Needle tip};
			
			\draw[white, thick, ->] 
			(1.7, 2.0) -- (2,1.6);
			\node[white, font=\bfseries] at (1.7, 2.2) {Lesion};
			
		\end{tikzpicture}
		
	\end{minipage}
	\subcaption{}
	\label{fig:phantom_diagram_c}
\end{subfigure}
	
	\caption{(a) Diagram of the \textit{ex vivo} tissue phantoms used for the usability studies.
		Dark purple lesions were present in the phantoms used for both studies, while light purple lesions were only present in the phantom used for the first study.
		(b) Photograph of the phantom in-use during the usability tests. (c) Screenshot of the tracking software taken during the usability study, annotated (white) to show the locations of the target lesion and needle tip.}
	\label{fig:phantom_diagram}
\end{figure}

Both studies used adapted versions of the \textit{ex vivo} tissue phantom used for the \textit{ex vivo} tissue accuracy study.
To simulate focal lesions undergoing biopsy, 2.5\,cm $\times$ 2.5\,cm $\times$ 2.5\,cm cubes of cooked beetroot were placed between the layers of chicken breast.
A layer of bovine tissue (sirloin steak) was placed at the top of the phantom to provide a flat and continuous surface against which the clinician would place the tracking and imaging probe assembly.
Ultrasound coupling gel was applied to the surface of the bovine tissue before covering it with 10\,\textmu m cling film to hold it in place.
The phantom used for the first study comprised six beetroot inclusions at three different depths, as shown in \autoref{fig:phantom_diagram}.
However, analysis of the data collected during the first study indicated that the varying inclusion depths introduced a confounding factor, potentially obscuring any true difference in procedure duration between tracked and untracked insertions.
To mitigate this, the phantom used for the second study was simplified to include only two simulated lesions, both positioned at 6\,cm.

Each clinician was given a few minutes at the beginning of their session to familiarise themselves with the phantom, probe assembly and tracking interface; during this time, the clinicians inserted the needle into the phantom under ultrasound guidance, but no biopsies were taken and no data was recorded.
Clinicians were then asked to perform at least six biopsy procedures; however, in the second study, one participant (a specialist registrar) was only able to complete 5 procedures in the time available, one participant (fourth-year trainee) completed 11 insertions and one participant (sixth-year trainee) completed 7 insertions.
Each procedure began with the insertion of the introducer needle and stylet to the target lesion under ultrasound guidance. The stylet was then removed, leaving the introducer needle in place, through which the biopsy needle was advanced into the lesion. A core needle biopsy was performed, the biopsy needle was withdrawn, and the sampled material was subsequently examined.
For the first study, inclusions at all three depths were targeted twice: once with tracking enabled and once without.
Clinicians could choose freely whether to target lesions of the left-hand or right-hand side of the phantom.
For the second study, clinicians carried out three tracked and three untracked insertions, and were free to choose which of the two~6\,cm deep inclusions to target for each.
The order of the insertions (depth and tracked/untracked for study one, and only tracked/untracked for study two) was randomised using a Python script executed prior to the arrival of each clinician.
For each insertion, the time between the puncture of the film and the clinician's verbal confirmation that the introducer needle had reached the intended target was recorded.
The success of the procedure was also recorded: if the extracted material appeared purple in colour, it was assumed to have come from a beetroot inclusion and the procedure was logged as successful.

After each session, the participant was sent an online questionnaire about their experience during the study; these were usually completed within a few days.
The questionnaire asked participants their agreement on a Likert scale with statements in four categories: device hardware, ultrasound imaging, tracking visualisation and tracking performance.
Free-text fields were provided for additional comments related to each category.
 A final statement was presented about the overall impact of the tracking technology on the ease of the procedure.
There were two further free-text fields: one for a description of the clinicians' previous experience with ultrasound-guided percutaneous needle procedures, and one for general comments about their experience with the technology.
Statements were worded such that agreement with each indicated satisfaction with the element of the device or process under consideration.

\section{Results}
\label{sec:results}
\subsection{Tracking Accuracy in Water}

\begin{figure}[h!]
\centering
\includegraphics[width=0.9\textwidth]{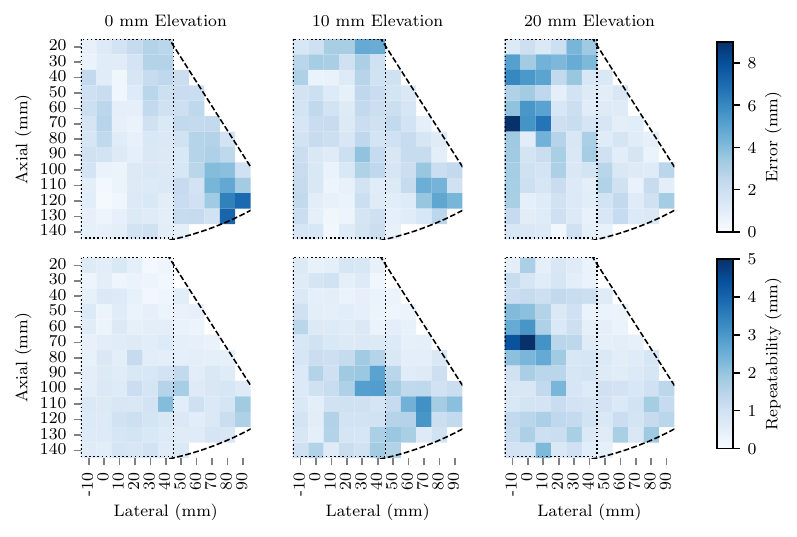}
\caption{Results of the measurement of tracking accuracy in water.
	The top row presents the magnitude of the 3D tracking error in the three elevational planes tested.
	The bottom row presents the standard deviation of this error (i.e. the tracking repeatability).
	The dashed line indicates the extent of the field of view of the ultrasound imaging system.
	The dotted line indicates the region of interest within which accuracy was further analysed.}
\label{fig:water_results}
\end{figure}

The results of the measurement of tracking accuracy in water up to a depth of 14\,cm are presented in \autoref{fig:water_results}.
In water, tracking accuracy was assessed: over the whole volume; within each of the three elevational planes tested; and in a region-of-interest (ROI) within each plane corresponding to the central part of the field-of-view of the ultrasound imaging system, where the needle tip is likely to reside throughout a procedure.
Average results across each plane and within the ROI in each plane are presented in \autoref{tab:water_results}.
With the needle tip in-plane (i.e. at 0\,mm elevation), the tracking system achieved an average error of 1.8\,mm and an average repeatability of less than 0.7\,mm, fairly uniformly over the whole plane; this error is close to the diameter of the introducer needle (1.47\,mm) and therefore deemed sufficient for accurately guiding the needle.
The spatial average tracking error magnitude over the whole volume was 1.96 $\pm$ 1.24\,mm, and the average repeatability was 0.95 $\pm$ 0.64\,mm.
In-plane, and within the ROI, average tracking error was 1.35 $\pm$ 0.68\,mm and average repeatability was 0.63 $\pm$ 0.31\,mm.

\begin{table}[h]
\centering
\caption{Spatial-average tracking error and repeatability magnitudes at different elevational positions, measured in water.
The extent of the region of interest is shown in \autoref{fig:water_results}.
Uncertainties are standard deviations.}
\label{tab:water_results}
\begin{tabular}{l c c c c}
\toprule
& \multicolumn{2}{c}{\textbf{Whole plane (mm)}} & \multicolumn{2}{c}{\textbf{Region of interest (mm)}} \\
\textbf{Elevation} & \textbf{Error} & \textbf{Repeatability} & \textbf{Error} & \textbf{Repeatability} \\
\midrule
0\,mm & $1.80 \pm 1.20$ & $0.66 \pm 0.33$ & $1.35 \pm 0.68$ & $0.63 \pm 0.31$ \\
10\,mm & $1.88 \pm 0.95$ & $1.02 \pm 0.62$ &  $1.80 \pm 0.87$ &  $0.95 \pm 0.55$ \\
20\,mm & $2.21 \pm 1.48$ & $1.16 \pm 0.76$ & $2.54 \pm 1.63$ &  $1.30 \pm 0.85$ \\
\textit{All planes} & $1.96 \pm 1.24$ & $0.95 \pm 0.64$ & $1.90 \pm 1.24$ & $0.96 \pm 0.67$ \\
\bottomrule
\end{tabular}
\end{table}

\subsection{Tracking Accuracy in an \textit{ex vivo} Tissue Phantom}
\begin{figure}[h!]
    \centering
    \begin{subfigure}[b]{0.45\textwidth}
        \centering
        \includegraphics{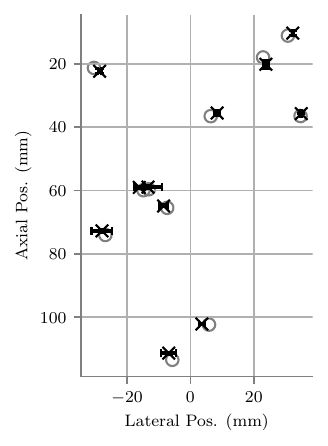}
        \subcaption{}
        \label{fig:xct_axial_lateral}
    \end{subfigure}%
    \begin{subfigure}[b]{0.3\textwidth}
        \centering
        \includegraphics{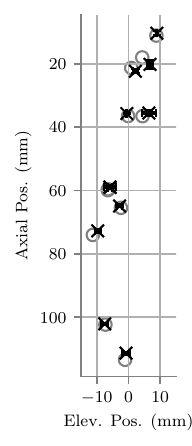}
        \subcaption{}
        \label{fig:xct_axial_elevational}
    \end{subfigure}
    \caption{Comparison of needle tip locations determined by X-ray CT (circles) and the tracking device (crosses).
    	The tracked positions are shown with error bars representing two standard deviations of the 100 measurements of tracked position made at each location, which at many locations are too small for effective visualisation.
    	Positions are plotted in the (a) lateral-axial and (b) elevational-axial planes.}
    \label{fig:xct_results}
\end{figure}

\autoref{fig:xct_results} compares tracked needle positions in the \textit{ex vivo} tissue phantom with tip locations derived using CT. The means and standard deviations of the repeat measurements of tracked position are plotted.
The spatial-average of the distance between tracked and
CT-derived positions was 2.04 $\pm$ 0.8\,mm.
The largest distance was 3.02 $\pm$ 1.24\,mm which occurred at a depth of 36.5\,mm according to CT.
Measurements made at a depth of 59.59\,mm had the poorest repeatability in tracked position, with a distance between tracked and true of 0.96 $\pm$ 2.4\,mm.

\subsection{Simulation Study of the Effects of Sound Speed Heterogeneity}

\begin{figure}
	\centering
	\includegraphics[width=0.8\textwidth]{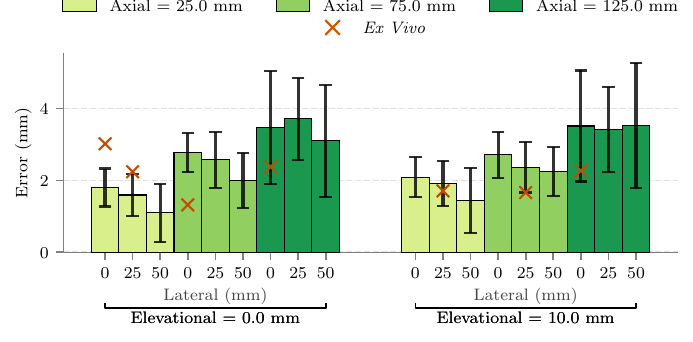}
	\caption{Tracking accuracy obtained from k-Wave simulations with heterogeneous sound-speed distributions, compared with \textit{ex vivo} results. Plotted simulated results are mean values, with error bars indicating the standard deviations. Orange crosses show average errors encountered during the \textit{ex vivo} study in the same spatial regions, where available: \textit{ex vivo} results were binned according to the their nearest simulated location and then averaged.}
	\label{fig:simulation_bar}
\end{figure}

The mean and standard deviations of the simulated tracking error are shown in \autoref{fig:simulation_bar}, with errors calculated relative to the true simulated needle locations and averaged across the ten digital phantoms.
Average tracking error increased from 1--2\,mm at a depth of 25\,mm  to 3--4\,mm at a depth of 125\,mm.
The results are plotted in comparison with the average error encountered during the \textit{ex vivo} tissue study in the same spatial regions, where available.
 
\subsection{\textup{Ex vivo} Tissue Phantom Demonstration}

\begin{figure}[h]
	\begin{subfigure}{0.32\textwidth}
		
		\begin{tikzpicture}
			\node[anchor=south west, inner sep=0] (img) 
			{	\includegraphics[trim=20cm 15cm 18.5cm 4cm, clip, width=\textwidth]{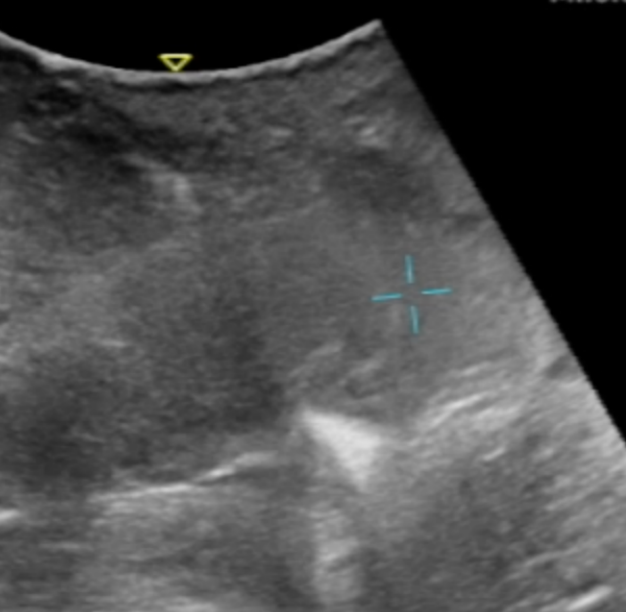}};
			\draw[red, dotted, thick] (2.7, 1.3) circle (14pt);
			\node[white, font=\bfseries] at (2.7,0.6) {Lesion};
		
		\end{tikzpicture}
		
	\subcaption{}
	\end{subfigure}
	\begin{subfigure}{0.32\textwidth}
		\includegraphics[trim=20cm 15cm 18.5cm 4cm, clip, width=\textwidth]{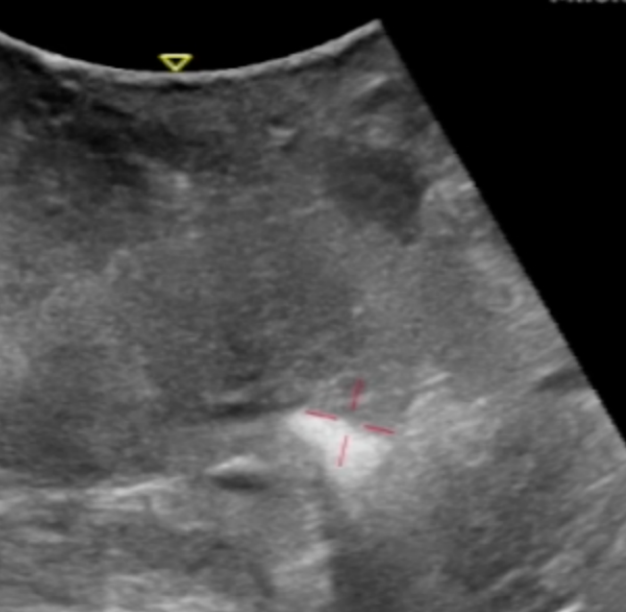}
		\subcaption{}
	\end{subfigure}
	\begin{subfigure}{0.32\textwidth}
		
				\begin{tikzpicture}
			\node[anchor=south west, inner sep=0] (img) 
			{	\includegraphics[trim=20cm 15cm 18.5cm 4cm, clip, width=\textwidth]{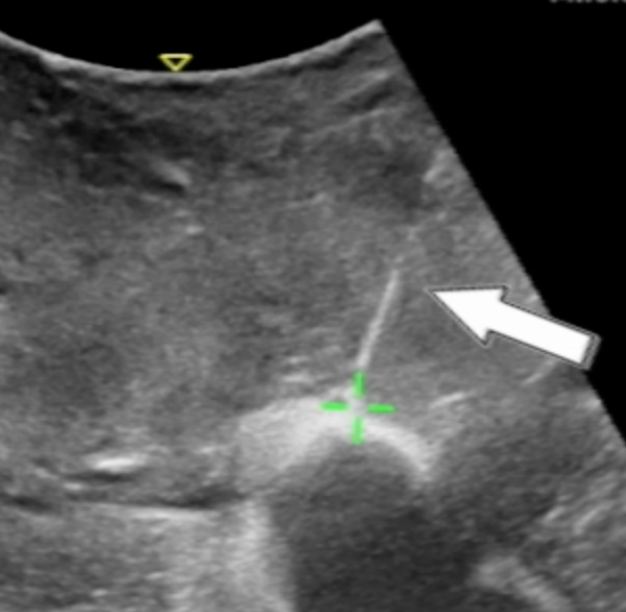}};
			\node[white, font=\bfseries, align=center] at (4.2,1.4) {Needle\\shaft};
			
		\end{tikzpicture}

		\subcaption{}
	\end{subfigure}
\caption{Cropped frames from the recorded video of the \textit{ex vivo} phantom demonstration. (a) shows the needle tip on its approach to the lesion (which is annotated with a red dashed circle); the needle tip is outside of the imaging plane, as indicated by the colour and tilt of the tracking cursor. (b) shows the needle tip once it has reached the target lesion, while out-of-plane. (c) shows the needle tip at the surface of the lesion, after the orientation of the imaging probe has been adjusted to place the needle tip and lesion in-plane. The location of the needle shaft is annotated in (c). The full video is available as a supplementary file.}
\label{fig:demo}
\end{figure}

\autoref{fig:demo} presents frames from the recorded video of the \textit{ex vivo} tissue phantom demonstration.
The full video is available as a supplementary file.
The video demonstrates how the 3D tracking information aids with the correct orientation of the needle and probe for the approach to the lesion, and how tracking functions even when the needle is not visible within the ultrasound image.
The tracking rate was 10.5\,Hz.

\subsection{Usability Tests in an \textup{ex vivo} Tissue Phantom}

\begin{figure}[htb!]
	\centering
	\begin{subfigure}[b]{0.75\textwidth}
		\centering
		\includegraphics[width=\textwidth]{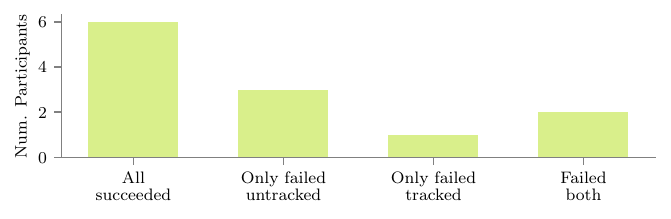}
		\caption{}
		\label{fig:contingency}
	\end{subfigure}
	\hfill
	\begin{subfigure}[b]{0.9\textwidth}
		\centering
		\includegraphics[width=\textwidth]{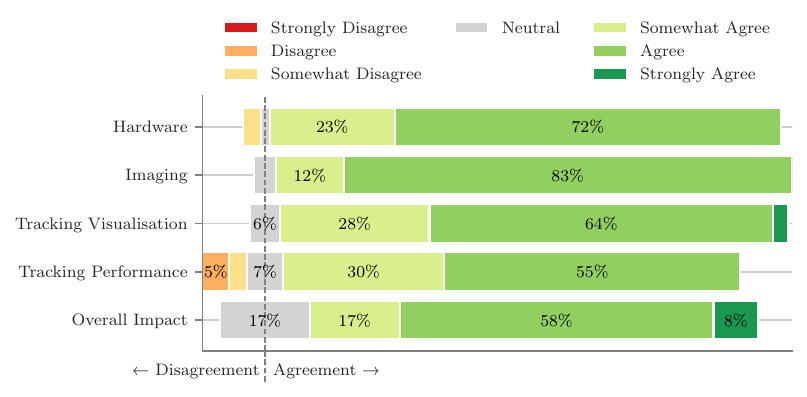}
		\caption{}
		\label{fig:survey_results}
	\end{subfigure}
	\caption{Usability test results. (a) Number of usability participants who succeeded in all procedures or failed at least one, with and without tracking. (b) Survey results aggregated into categories. Statements were worded such that agreement indicates satisfaction with the element of the device or process under consideration. Annotations on each bar indicate the percentage of responses in each category at each level of agreement. Percentages less than 5\% are not annotated.}
	\label{fig:combined_usability}
\end{figure}

The biopsy failure rate was 15.8\% with tracking disabled and 10.3\% with tracking enabled (a 35\% reduction).
A Fisher's Exact Test, chosen because of the small number of samples and low overall number of failures, yielded a p-value of 0.5.
For junior doctors (i.e. non-consultants), the failure rate was 19.2\% with tracking disabled and 14.8\% with tracking enabled (a 23\% reduction).
For consultants, the failure rate was 8.3\% without tracking, while no consultants failed any tracked procedures.
\autoref{fig:contingency} presents the number of participants who: succeeded in all biopsies; failed only untracked biopsies; failed only tracked biopsies; and failed at least one biopsy in both categories.
The six most experienced participants (out of 12) succeeded in all procedures both with and without tracking.
Three participants (one consultant and the two least-experienced trainees) succeeded in all tracked procedures, but failed in at least one untracked procedure.
Only one participant, a sixth-year trainee, failed a tracked procedure while succeeding in all untracked procedures.
Two participants, both moderately experienced, failed at least one procedure in both conditions.
Tracked procedures took, on average, 37 $\pm$ 25\,s. This was slightly longer than untracked procedures, which took 31 $\pm$ 32\,s.
A Mann-Whitney U Test, chosen because of right-skewed distribution of procedure duration, yielded a p-value of 0.07.

Aggregated results of the Likert survey are shown in \autoref{fig:survey_results}. Individual results are tabulated in the supplementary materials.
The aggregated results show satisfaction across all categories.
However, two participants felt that the presence of the tracking array on the imaging probe interfered with their ability to carry out the procedure.
No dissatisfaction was reported with ultrasound imaging or the tracking cursor graphic.
There were five negative responses in the tracking performance category: four indicating that tracking performance was affected by depth, and one participant who perceived some latency in the tracking visualisation.
All participants felt tracking was accurate enough to support the procedure.
Ten participants found procedures easier with tracking enabled, with the remaining two participants reporting no benefit.

In the free-text fields of the survey, comments from clinicians were broadly positive.
Participants described the device as a ``very good idea'' and a ``useful concept'' with the potential to make procedures ``more accurate and safe.''
The value of the device in specific, challenging scenarios was highlighted, with clinicians expecting benefits when ``targeting deeper and harder to visualise lesions'' or ``in the presence of excessive fat [or] air''.
Several participants commented that the system would be an excellent training tool for less experienced clinicians.
While the clinicians found the trackable stylet fitted well with the introducer needle, a recurring theme in the hardware category was the ``bulky'' and heavy nature of the tracking array mounted on the ultrasound probe.
Some feedback from the first study suggested improving the colour map to make the cursor more clearly visible when out of plane.
The updated colour map used for the second study received positive feedback.
Some clinicians raised concerns about accuracy degradation at greater needle depths, where occasional ``jumping'' of the tracking cursor was reported during the first study.

\section{Discussion}
\label{sec:discussion}
Our solution is unique amongst competing ultrasonic needle tracking techniques in providing truly quantitative, 3D positional information through the use of MLAT and a 2D sparse array. This is in contrast to existing methods which quantify in-plane position---using a dense, 1D imaging array---and then estimate the elevational position from the properties of the received signals~\cite{bakerIntraoperativeNeedleTip2022, kasineNeedleTipTracking2019, deepsighttechnologyinc.DeepSightTechnologyReceives, tanakaSplitbasedElevationalLocalization2024}.
MLAT has been widely deployed in global positioning (GPS), aerospace and indoor localisation \cite{munozPositionLocationTechniques2009}.
The use of MLAT for localising an acoustic source or receiver requires far fewer elements than methods based on image reconstruction.
Ultrasonic MLAT has previously been employed for tracking a catheter with an integrated 3.5\,MHz piezoelectric radial transmitter and a 3 $\times$ 3, 15\,cm $\times$ 20\,cm receiver array; the receiver array was integrated into an immobiliser for a swine model within which catheter tracking was tested for stereotactic endovascular aortic navigation \cite{mungStereotacticEndovascularAortic2013}.
Similarly, ultrasonic MLAT has been applied to tracking of a photoacoustic ultrasound transmitter placed pre-operatively within a simulated lesion in a cadaveric human breast, to assist with breast conservation surgery; the location of the transmitter was tracked using a triangular 3-element array with an element spacing of approximately 20\,mm \cite{lanFiberOptoacousticGuide2018}.
MLAT is commonly implemented using (weighted) least-squares optimisation, which is equivalent to MLE under Gaussian measurement noise.
Our MLAT implementation employs a full MLE formulation, enabling the use of measurement-specific, non-Gaussian range uncertainty models that are generated dynamically from the amplitudes of the received signals (see \autoref{sec:software} and supplementary materials), accurately modelling and weighting each measurement according to its inferred uncertainty.

Exploiting a needle-integrated transmitter, rather than receiver, allows the needle location to be resolved from a single ultrasonic transmission, increasing the maximum tracking frame rate.
Feedback from usability study participants revealed that the provision of quantitative, high-confidence 3D tracking information greatly improved the clinicians' abilities to generate and navigate the 3D mental maps of the relative locations of the imaging probe, lesion, critical structures and needle tip required to safely and effectively manoeuvre the needle to the target, reducing cognitive burden.
While the current prototype was developed with focal liver lesion biopsy as the intended use, the technology is applicable to a wide range of ultrasound-guided needle procedures including other solid organ procedures such as fiducial marker placement and percutaneous ablation therapy, as well as procedures for fetal medicine and regional anaesthesia.
The transmitter could also be readily integrated into catheters for vascular access and other interventional radiology procedures, which may require registration with video fluoroscopy or other non-ultrasound imaging modalities.

\subsection{In-plane Tracking Accuracy}
Within the in-plane ROI, tracking error and repeatability was about twice that of our previous, 2D (in-plane only) tracking system, which tracked the needle tip by analysing B-mode ultrasound imaging pulses received by a needle-integrated fibre-optic hydrophone, and achieved 0.67\,mm and 0.28\,mm spatial-average error and repeatability, respectively, in water, over a comparable field of view~\cite{bakerIntraoperativeNeedleTip2022}.
Similar accuracies have been reported for other 2D, receive-mode methodologies \cite{deepsighttechnologyinc.DeepSightTechnologyReceives, kasineNeedleTipTracking2019}.
The reduced tracking signal frequency of our 3D system (1\,MHz) compared to our previous 2D system (3\,MHz) and other similar technologies may account for this decreased tracking resolution.
The lower frequency was necessary to provide a wide directivity to the elements of the sparse tracking array, which, because of its low element count, was not capable of beam-forming, in contrast to the curvilinear imaging array used for the 2D methodologies.
The low-element-count sparse array, however, is what enabled cost-effective, quantitative 3D tracking.

The in-plane accuracy was affected by a small region of large errors in the bottom-right corner of the field of view.
The errors are also seen to increase with distance from the ultrasound probe in the regions above.
However, the good repeatability in this region indicated this was not random error, but a systematic effect.
While the larger error was present in both independent repeats of the experiment, it was much more pronounced in one repeat (9\,mm) compared to the other (5\,mm).
Rotational misalignment of the array assembly and the grid of ``true'' locations would have the greatest effect in this region, right in the corner of the field-of-view.
The error was also in different directions for each experiment, suggesting its source is unlikely to be intrinsic to the tracking system.
Although these regions are theoretically more susceptible to errors due to the reduced subtended angle and resulting Geometric Dilution of Precision (GDOP), examination of the tracking waveforms exhibited a strong signal-to-noise ratio.
No artifacts, such as noise peaks mistaken for tracking pulses, were observed.
It is likely that the primary cause is experimental error during alignment of the probe assembly and motion control system used to derive true locations.

\subsection{Out-of-plane Tracking Accuracy}
Tracking accuracy was assessed up to 20\,mm from the imaging plane.
While the system is capable of tracking at larger elevational distances from the plane, consultation with clinicians during the design process identified a tracking volume spanning $\pm$ 20\,mm from the imaging plane was suitable and provided the tracking visualisation (\autoref{fig:combined_cursor_figures}) with a useful amount of sensitivity.
In the plane 10\,mm from the imaging plane, repeatability and error remained close to the diameter of the needle.
In the plane 20\,mm from the imaging plane, results were similar with the exception of a poor accuracy region around the centre of the field of view and shallow axial locations, where accuracy peaked at 9\,mm with a corresponding peak in repeatability of 5\,mm.
The recently described quantitative 3D tracking method based on a fibre-optic ultrasound transmitter and 1D imaging array, which estimated elevational location from the morphology of the received signals, demonstrated tracking beyond $\pm$ 20\,mm but only at a depth of 26\,mm; the authors anticipate that at a depth of $\pm$ 100\,mm tracking would only be possible up to $\pm$ 5\,mm~\cite{tanakaSplitbasedElevationalLocalization2024}.

In these planes, spatial variation in error was consistent with that of repeatability, indicating random error rather than any systematic effects.
The peak in error in the 20\,mm plane also corresponded with a disagreement between the two repeat experiments, suggesting that experimental error may have been a contributing factor.
The mean difference in measured tracking error between the two repeated experiments, averaged over all spatial locations, was 1.29 $\pm$ 0.87\,mm, indicating good agreement between the two experiments overall.
The accuracy of the alignment procedure relied on the quality of the image of the needle generated by the ultrasound imaging system, as well as the judgment of the user to correctly position the needle in the centre of the imaging plane, with its tip on the cross-hair shown in \autoref{fig:combined_cursor_figures}.
These same factors also introduce uncertainty into the registration of the tracking array and imaging system, which was carried out independently to the water accuracy experiments but comprised a similar procedure.

The increase in repeatability as the needle moved away from the imaging plane is due to the reduction in signal amplitude at the array elements on the opposite side of the imaging plane, due both to the increased propagation distance and the directivity of the receiving elements; the spatial-average repeatability was, however, smaller than the needle diameter in all planes.
Over the whole volume, tracking error was 1.96\,mm with a repeatability of 1.24\,mm on average, which is close to the diameter of the needle and therefore deemed suitable for accurate needle guidance.
The waveform-morphology 3D tracking method had a reported accuracy of 2.4\,mm, despite the use of a high-frequency imaging probe (5--11\,MHz) which limited experiments to a depth of 26\,mm; while tracking is possible at larger depths, both the range of trackable elevational locations and the tracking accuracy are reduced~\cite{tanakaSplitbasedElevationalLocalization2024}.
This compares to an accuracy of 0.4\,mm achieved by our previous 3D tracking method which exploited a needle-integrated hydrophone, a research-mode ultrasound system and a hybrid imaging-based / MLAT tracking approach, but could only track up to 15\,mm from the imaging plane and to depths of 38\,mm due to its high-frequency transducers (4--9\,MHz)~\cite{xiaLookingImagingPlane2017}.
Our new method is independent of ultrasound imaging, and therefore compatible with clinical ultrasound imaging systems.

\subsection{Registration and Alignment}
Both registration of the tracking array to the imaging probe and the alignment completed for accuracy assessment in water required measurement of true locations of the needle tip relative to the imaging probe.
Determining these locations without relying on a second tracking technology, which itself would require validation and registration, posed a significant challenge.
The use of the ultrasound imaging system itself to localise the needle tip in water, where contrast is high, provided an intrinsic solution.
However, the ultrasound slice thickness and image resolution limited the precision with which it was possible to place the needle tip in-plane and at the vertices of the cross-hair.
This impacted the results of the water accuracy experiment, particularly at the edges of the domain where rotational misalignment contributes as greater error, as discussed above, while any error in registration will degrade tracking accuracy.
In future, alignment and registration, which takes place in water where there is good contrast and homogenous sound speed, could be improved by employing automatic analysis of the ultrasound imaging feed to record pixel intensity around the needle tip location and determine when it was in-plane.
For the purposes of alignment, a closed-loop system would automatically move the needle until the tip was in the desired location in the ultrasound image.
For registration, it may be sufficient to manually manipulate the position of the needle beneath the probe assembly while performing this analysis, continuously comparing the tracked locations with the apparent locations of the needle tip within the ultrasound image.
The results would, however, be dependent on the performance of the ultrasound image processing algorithm, which itself would need validation.

\subsection{Needle Orientation}
It should be noted that water measurements of tracking error were carried out with the needle pointing horizontally, parallel to the lateral axis of the imaging system, and that alignment was carried out by localising the tip of the needle with the ultrasound image.
The precise location of the fibre-optic ultrasound transmitter within the needle bevel, however, was not accounted for.
The bevel of the 18 gauge trackable stylet was approximately 6\,mm long.
Since the system was registered using the same trackable needle at the same orientation as used for accuracy measurement, this error will have been accounted for by registration.
However, during clinical use, where the orientation of the needle and therefore the direction of this offset varies, transmitter location within the needle tip may contribute an additional uncertainty.
While directivity of the needle-integrated transmitters was not measured for this study, little to no degradation in tracking accuracy with insertion angle has been observed, so long as the needle bevel remains facing upwards, towards the probe assembly.
This is evidenced by the \textit{ex vivo} tissue accuracy results, for which steep insertion angles were used.
Our previous work with similar needle-integrated transmitters demonstrated their omnidirectional  ultrasound emission~\cite{xiaUltrasonicNeedleTracking2017}.

For the determination of needle tip position from CT images, the main contribution to uncertainty was in the manual estimation of the orientation of the imaging probe relative to the CT coordinate system.
Empirical assessment suggested a confidence in the estimated orientation of approximately 1$^\circ$, which, at a depth of 120\,mm, would result in an error in $p_{\textrm{CT}}$ of 2\,mm in the direction of the rotational error.
The relatively low standard deviations in tracked location compared to the error magnitudes support the hypothesis that there is a strong systematic component such as this to the tracking error in the phantom.

\subsection{Sound Speed Heterogeneity}
The tracking algorithm assumed a homogeneous speed of sound of 1540\,m/s in tissue and 1480\ m/s in water.
The speed of sound in the \textit{ex vivo} tissue accuracy phantom likely varied between 1480\,m/s (in pockets of water between the chicken tissue) and up to 1600\,m/s \cite{shishitaniChangeAcousticImpedance2012}.
This heterogeneity will have introduced some bias to the MLAT due to the invalidity of the conversion of time-of-arrival to range and bending of acoustic paths due to refraction.
These errors will manifest when tracking is compared to non-acoustic imaging modalities such as CT, but in practice, when ultrasound imaging is used for guidance, tissue anatomical information and the location of the procedure target within the ultrasound image will have been affected by the heterogeneity in a similar way, reducing the clinical impact of such errors.
Our system achieved an average tracking error of 2.05\,mm in the \textit{ex vivo} phantom.
Our simulation study predicted an average tracking error of 2.5 $\pm$ 1.0\,mm within the heterogeneous anatomy expected to be encountered during a clinical procedure, while a previous simulation study found that a 5\% speed of sound heterogeneity introduced tracking errors of 5\,mm on average, when 3D tracking was performed by combining analysis of the received signal amplitudes for in-plane localisation with MLAT for elevational localisation, at depths up to 60\,mm~\cite{xiaLookingImagingPlane2017}.

While our algorithm currently assumes a uniform speed of sound, it may be possible to jointly estimate spatially-varying speed of sound within the MLE, providing a speed of sound map and improved tracking in heterogeneous media.
Previously, deep learning has been applied to the estimation of sound speed from 2D ultrasound images \cite{shiLearningbasedSoundSpeed2024} which may assist with a MLE estimation of sound speed, although the extrapolation of such estimates into 3D would pose a challenge.
Importantly, such joint estimation would only be meaningful if the resulting heterogenous speed-of-sound map was also incorporated into the ultrasound image reconstruction process.
Otherwise, inconsistencies would arise between the tracked needle position and the imaged anatomy, as conventional ultrasound imaging assumes a homogenous sound speed.
Joint estimation would also be significantly more computationally intensive.
For these reasons, such a method was not investigated for the current prototype.

\subsection{Usability Study}
Half of participants in the usability tests---the six most experienced clinicians---succeeded in all performed biopsies, with and without tracking.
Overall, the biopsy failure rate was only modestly reduced by the use of tracking, and the difference between the two groups was not statistically significant.
This suggests that the phantom biopsies did not represent the clinical situations for which ultrasound-guidance is most challenging.
Improving the phantom by adding some inclusions representing critical structures that must be avoided could better demonstrate the impact of tracking on failure rates.
Notably, only one clinician succeeded in all untracked procedures but failed a tracked procedure.
Similarly, three clinicians failed any tracked procedure, while five clinicians failed at least one untracked procedure.
When combined with the positive qualitative results from the survey, and individual feedback from clinicians, these findings suggest that the tracking system made the procedures easier to carry out.

The over-simplicity of the simulated biopsies was also reflected in the procedure duration results, which failed to demonstrate a statistically significant difference between tracked and untracked procedures.
In fact, on average, tracked procedures took slightly longer to complete.
It was observed that, despite being given time to practice beforehand, clinicians were initially hesitant during tracked insertions, being seen to experiment with the tracking visualisation by moving the ultrasound probe relative to the needle, in part to gauge its sensitivity to movement.
Some clinicians also needed reminding of the meaning of the colour coding and orientation during insertions, increasing durations.
While we anticipate that comprehensive training and routine use of the tracking device would result in shorter biopsies (and likely further reduced failure rates), this duration of the learning curve could likely be reduced by the addition of on-screen information about the meaning of the appearance of the tracking graphic; a colour bar, for example.
An additional 3D graphic showing the location of the needle tip relative to the imaging plane may also help reduce the learning curve.
However, even without these improvements, we anticipate that comprehensive training and routine use of the tracking device would result in shorter biopsies and further reduced failure rates.

Overall, the results of the survey demonstrated that the participants considered that the device made the procedures easier to carry out.
Participants were satisfied with the hardware components of the device, although acknowledged that the bulky and heavy nature of the current tracking array prototype may make some insertions more difficult, particularly when it is necessary to tilt the probe assembly in the elevational direction when coupled to the patient, which may lift one side of the tracking array away from the patient, losing coupling.
A clinical prototype would likely require that the probe assembly have a smaller footprint, allowing tilting to take place without losing coupling; this may, however, reduce tracking accuracy in the elevational direction, which benefits from a wider elevational separation of tracking elements.
Participants were also satisfied with the imaging quality, latency and rate displayed from the imaging system via the frame grabber.
The small number of negative responses relating to tracking performance were mostly related to a perceived degradation of tracking performance with depth.
Deeper into the phantom, acoustic attenuation in tissue and the natural widening of the ultrasound field both act to reduce the signal-to-noise ratio at the tracking array, and therefore the accuracy of range measurements.
While the tracking accuracy results in the \textit{ex vivo} tissue phantom indicate that performance is sufficient at depth, the less well controlled environment of the usability study may have resulted in reduced tracking performance due to factors such as needle orientation (while clinicians were asked to keep the bevel facing up, this was not always adhered to) and degradation in the quality of the phantom and trackable needles during repeated use.

\subsection{Transmitter Durability}
To reduce risk of infection, focal liver lesion biopsy needles are required to be single use, and our low-cost design is well suited to this (see \autoref{sec:clinical}).
While this means that during clinical practice, transmitters will only need to last for the duration of a single procedure, it was observed during the usability study that the acoustic output of the transmitters sometimes reduced during the multi-procedure session.
During the water accuracy measurements, transmitter performance was found to remain stable for more than 24h of constant excitation in water, suggesting that the performances changes seen during the usability study were likely due to mechanical factors such as compressed tissue or air filling the needle bevel and obstructing the generation of ultrasound, further evidenced by the recovery of the performance of some transmitters after cleaning.
Encapsulation of the transmitter within the bevel with a thin layer of protective material such as Parylene C or additional PDMS may prevent this.
Future work will quantify the lifetime of the devices in water and with repeated insertions in tissue.

\subsection{Clinical Translation}
\label{sec:clinical}
Development of a clinical prototype will focus on two aspects: improvements to the ergonomics of the probe assembly (including reducing its footprint) and improvements to the fibre-optic ultrasound transmitters.
Methods to increase the acoustic output of the transmitters at 1\,MHz (the centre frequency of the tracking array) are currently being investigated.
This may allow truly simultaneous imaging and tracking (rather than the synchronisation of tracking measurements between imaging frames), for which tracking performance is currently degraded due to acoustic backscatter from the imaging frames.
Simultaneous imaging and tracking would also be subject to interference from tracking signals on the imaging frame, and methods to reduce the output of the transmitters within the imaging band are also currently being investigated, including the use of a continuous wave laser for band-limited photoacoustic generation~\cite{allesAdaptiveLightModulation2016}.
Simultaneous tracking and imaging would enable faster imaging and tracking frame rates, and remove the requirement for a trigger signal connection between the imaging and tracking systems.

We anticipate that mass produced, single use trackable needles for a future clinical system will be competitively priced against equivalent solid needle stylets, at around £10.
The dip coating and assembly procedure could be easily automated for mass production.
For the current prototype, the substantial majority of the component cost comes from the precision SMA905 optical connector, whereas a clinical device would use simple, low-cost plastic connectors, reducing this to under £1.
The remaining material costs are already minimal and would become negligible in mass production.
This is in contrast to existing trackable needle technologies featuring needle-integrated EM~\cite{hakimeElectromagneticTrackedBiopsyUltrasound2012}, piezoelectric~\cite{kasineNeedleTipTracking2019} or fibre-optic~\cite{deepsighttechnologyinc.DeepSightTechnologyReceives} sensors, where costs can be up to a few hundred pounds per needle.

Clinical acceptance of the technology will benefit from the minimisation of changes to the current clinical workflow.
It will therefore be important to improve the ergonomics of the hardware when developing a clinical prototype, and ensure that any clinical device is compatible with existing procedures.
We envisage that a complete interventional ultrasound system with 3D tracking capability would provide this.
Real-time visualisation of the 3D location of the needle tip during ultrasound-guided procedures, even when the tip is not visible within the ultrasound image, would enable clinicians to confidently, precisely and safely navigate the route to the clinical target, reducing the cognitive burden associated with the 3D mental mapping that is necessary to complete these procedures under 2D ultrasound guidance.
Technologies that enable easy and precise needle insertions would also reduce trauma and improve outcomes, and support more junior staff in performing procedures that would otherwise have required the presence of a senior clinician.
Facilitating the safe, fast, and precise guidance of complex procedures using ultrasound, rather than more expensive and resource-intensive imaging modalities like CT, would reduce procedure durations and costs.
This would increase capacity and alleviate waiting times.
Accelerating training and reducing the learning curve for junior practitioners would also help to alleviate staffing shortages and expand service capacity \cite{RCRSurvey2023} during a time of ever-increasing cancer prevalence \cite{CancerIncidenceCommon2020}.

\section{Conclusion}
\label{sec:conclusion}
We have presented a quantitative 3D ultrasonic needle tracking system for ultrasound-guided focal liver lesion procedures, evaluating its accuracy in water and \textit{ex vivo} tissue phantoms, and assessing its usability with 12 clinicians.
The system comprised a needle-integrated fibre-optic ultrasound transmitter and a 16-element, sparse tracking array mounted to a curvilinear imaging probe.
In water, the system achieved sub-millimetre tracking repeatability within the imaging plane and an average tracking error of 1.96\,mm over a volume covering half of the field of view of the imaging probe to depths of 140\,mm and up to 20\,mm away from the imaging plane.
In the \textit{ex vivo} tissue phantom, the system maintained an average tracking error of 2.05\,mm, indicating resilience to sound-speed heterogeneity and acoustic attenuation in tissue.
Accuracy was found to be close to the diameter of the needle, and therefore suitable for needle guidance.

The usability study demonstrated that the system made procedures easier to perform, particularly for less-experienced clinicians, with positive survey feedback regarding hardware, imaging, tracking and visualisation.
The biopsy failure rate was reduced by 35\%, and it is believed that with a more realistic phantom or an \textit{in vivo} demonstration, which would present more challenging procedures, improvements in failure rates would be more significant.
While procedure durations were slightly increased by the use of tracking, this was highly likely to be due to the novelty of the tracking graphic rather than limitations of the tracking system.
Some challenges identified during the usability study included the size of the current probe assembly and depth-related performance degradation.

Future work will focus on improving the ergonomics of the probe assembly, increasing the acoustic output of the fibre-optic transmitters, and enabling truly simultaneous imaging and tracking.
These developments aim to further improve tracking accuracy and clinical usability.
Overall, the results demonstrate that the system provides accurate and reliable needle tracking, with the potential to improve procedural outcomes, particularly in challenging clinical scenarios or for training less-experienced operators, while remaining compatible with standard interventional ultrasound workflows.

\ack{The authors would like to thank India Lewis-Thompson at University College London for her guidance setting up a fibre-optic ultrasound transmitter manufacturing capability at King's College London. We would also like to thank the Interventional Radiologists from Guy's and St Thomas' NHS Trust and Barts Health NHS Trust who participated in the usability study.}

\funding{This work was supported by the Beijing Institute of Collaborative Innovation.}

%
\data{Data and Python scripts for generating Figures 5, 6 and 8, are available at \url{https://github.com/KCL-BMEIS/RT3DUS-Needle-Tracking-PA-Beacon-Data}.}
%

\printbibliography

@article{bakerIntraoperativeNeedleTip2022,
  title = {Intraoperative {{Needle Tip Tracking}} with an {{Integrated Fibre-Optic Ultrasound Sensor}}},
  author = {Baker, Christian and Xochicale, Miguel and Lin, Fang-Yu and Mathews, Sunish and Joubert, Francois and Shakir, Dzhoshkun I. and Miles, Richard and Mosse, Charles A. and Zhao, Tianrui and Liang, Weidong and Kunpalin, Yada and Dromey, Brian and Mistry, Talisa and Sebire, Neil J. and Zhang, Edward and Ourselin, Sebastien and Beard, Paul C. and David, Anna L. and Desjardins, Adrien E. and Vercauteren, Tom and Xia, Wenfeng},
  year = {2022},
  month = nov,
  journal = {Sensors},
  volume = {22},
  number = {23},
  pages = {9035},
  issn = {1424-8220},
  doi = {10.3390/s22239035},
  urldate = {2024-11-29},
  copyright = {https://creativecommons.org/licenses/by/4.0/},
  langid = {english}
}

@article{beigiEnhancementNeedleVisualization2021,
  title = {Enhancement of Needle Visualization and Localization in Ultrasound},
  author = {Beigi, Parmida and Salcudean, Septimiu E. and Ng, Gary C. and Rohling, Robert},
  year = {2021},
  month = jan,
  journal = {International Journal of Computer Assisted Radiology and Surgery},
  volume = {16},
  number = {1},
  pages = {169--178},
  issn = {1861-6410, 1861-6429},
  doi = {10.1007/s11548-020-02227-7},
  urldate = {2024-11-29},
  langid = {english}
}

@misc{CancerIncidenceCommon2020,
  title = {Cancer Incidence for Common Cancers},
  year = {2020},
  journal = {Cancer Research UK},
  urldate = {2024-11-29},
  howpublished = {https://www.cancerresearchuk.org/health-professional/cancer-statistics/incidence/common-cancers-compared},
  langid = {english}
}

@article{hakimeElectromagneticTrackedBiopsyUltrasound2012,
  title = {Electromagnetic-{{Tracked Biopsy}} under {{Ultrasound Guidance}}: {{Preliminary Results}}},
  shorttitle = {Electromagnetic-{{Tracked Biopsy}} under {{Ultrasound Guidance}}},
  author = {Hakime, Antoine and Deschamps, Frederic and De Carvalho, Enio Garcia Marques and Barah, Ali and Auperin, Anne and De Baere, Thierry},
  year = {2012},
  month = aug,
  journal = {CardioVascular and Interventional Radiology},
  volume = {35},
  number = {4},
  pages = {898--905},
  issn = {0174-1551, 1432-086X},
  doi = {10.1007/s00270-011-0278-8},
  urldate = {2024-11-29},
  copyright = {http://www.springer.com/tdm},
  langid = {english}
}

@article{kasineNeedleTipTracking2019,
  title = {Needle Tip Tracking for Ultrasound-guided Peripheral Nerve Block Procedures---{{An}} Observer Blinded, Randomised, Controlled, Crossover Study on a Phantom Model},
  author = {K{\aa}sine, Trine and Romundstad, Luis and Rosseland, Leiv Arne and Ullensvang, Kyrre and Fagerland, Morten Wang and Hol, Per Kristian and Kessler, Paul and Sauter, Axel Rudolf},
  year = {2019},
  month = sep,
  journal = {Acta Anaesthesiologica Scandinavica},
  volume = {63},
  number = {8},
  pages = {1055--1062},
  issn = {0001-5172, 1399-6576},
  doi = {10.1111/aas.13379},
  urldate = {2024-11-29},
  langid = {english}
}

@article{khalifaUtilityLiverBiopsy2020a,
  title = {The Utility of Liver Biopsy in 2020},
  author = {Khalifa, Ali and Rockey, Don C.},
  year = {2020},
  month = may,
  journal = {Current Opinion in Gastroenterology},
  volume = {36},
  number = {3},
  pages = {184--191},
  issn = {0267-1379, 1531-7056},
  doi = {10.1097/MOG.0000000000000621},
  urldate = {2025-04-23},
  langid = {english}
}

@article{lanFiberOptoacousticGuide2018,
  title = {A Fiber Optoacoustic Guide with Augmented Reality for Precision Breast-Conserving Surgery},
  author = {Lan, Lu and Xia, Yan and Li, Rui and Liu, Kaiming and Mai, Jieying and Medley, Jennifer Anne and {Obeng-Gyasi}, Samilia and Han, Linda K. and Wang, Pu and Cheng, Ji-Xin},
  year = {2018},
  month = may,
  journal = {Light: Science \& Applications},
  volume = {7},
  number = {1},
  pages = {2},
  issn = {2047-7538},
  doi = {10.1038/s41377-018-0006-0},
  urldate = {2025-07-02},
  langid = {english}
}

@article{langbergEchotransponderElectrodeCatheter1988,
  title = {The Echo-Transponder Electrode Catheter: {{A}} New Method for Mapping the Left Ventricle},
  shorttitle = {The Echo-Transponder Electrode Catheter},
  author = {Langberg, Jonathan J. and Franklin, Jay O. and Landzberg, Joel S. and Herre, John M. and Kee, Laura and Chin, Michael C. and Bharati, Saroja and Lev, Maurice and Himelman, Ronald B. and Schiller, Nelson B. and Griffin, Jerry C. and Scheinman, Melvin M.},
  year = {1988},
  month = jul,
  journal = {Journal of the American College of Cardiology},
  volume = {12},
  number = {1},
  pages = {218--223},
  issn = {07351097},
  doi = {10.1016/0735-1097(88)90377-4},
  urldate = {2024-11-29},
  copyright = {https://www.elsevier.com/tdm/userlicense/1.0/},
  langid = {english}
}

@article{ledijubellPhotoacousticbasedVisualServoing2018,
  title = {Photoacoustic-Based Visual Servoing of a Needle Tip},
  author = {Lediju Bell, Muyinatu A. and Shubert, Joshua},
  year = {2018},
  month = oct,
  journal = {Scientific Reports},
  volume = {8},
  number = {1},
  pages = {15519},
  issn = {2045-2322},
  doi = {10.1038/s41598-018-33931-9},
  urldate = {2024-11-29},
  langid = {english}
}

@article{mathewsUltrasonicNeedleTracking2022,
  title = {Ultrasonic {{Needle Tracking}} with {{Dynamic Electronic Focusing}}},
  author = {Mathews, Sunish J. and Shakir, Dzhoshkun I. and Mosse, Charles A. and Xia, Wenfeng and Zhang, Edward Z. and Beard, Paul C. and West, Simeon J. and David, Anna L. and Ourselin, Sebastien and Vercauteren, Tom and Desjardins, Adrien},
  year = {2022},
  month = mar,
  journal = {Ultrasound in Medicine \& Biology},
  volume = {48},
  number = {3},
  pages = {520--529},
  issn = {03015629},
  doi = {10.1016/j.ultrasmedbio.2021.11.008},
  urldate = {2024-11-29},
  langid = {english}
}

@article{mungStereotacticEndovascularAortic2013,
  title = {Stereotactic Endovascular Aortic Navigation with a Novel Ultrasonic-Based Three-Dimensional Localization System},
  author = {Mung, Jay C. and Huang, ShihYau Grace and Moos, John M. and Yen, Jesse T. and Weaver, Fred A.},
  year = {2013},
  month = jun,
  journal = {Journal of Vascular Surgery},
  volume = {57},
  number = {6},
  pages = {1637--1644},
  issn = {07415214},
  doi = {10.1016/j.jvs.2012.09.078},
  urldate = {2025-07-02},
  langid = {english}
}

@book{munozPositionLocationTechniques2009,
  title = {Position Location Techniques and Applications},
  author = {Mu{\~n}oz, David},
  year = {2009},
  publisher = {Academic Press},
  address = {Amsterdam Boston},
  isbn = {978-0-12-374353-4},
  langid = {english},
  lccn = {910.285}
}

@techreport{RCRSurvey2023,
  title = {Clinical {{Radiology Workforce Census}} 2023},
  author = {{The Royal College of Radiologists}},
  year = {2023}
}

@article{vernuccioNegativeBiopsyFocal2019a,
  title = {Negative {{Biopsy}} of {{Focal Hepatic Lesions}}: {{Decision Tree Model}} for {{Patient Management}}},
  shorttitle = {Negative {{Biopsy}} of {{Focal Hepatic Lesions}}},
  author = {Vernuccio, Federica and Rosenberg, Michael D. and Meyer, Mathias and Choudhury, Kingshuk R. and Nelson, Rendon C. and Marin, Daniele},
  year = {2019},
  month = mar,
  journal = {American Journal of Roentgenology},
  volume = {212},
  number = {3},
  pages = {677--685},
  issn = {0361-803X, 1546-3141},
  doi = {10.2214/AJR.18.20268},
  urldate = {2025-04-23},
  langid = {english}
}

@article{xiaLookingImagingPlane2017,
  title = {Looking beyond the Imaging Plane: {{3D}} Needle Tracking with a Linear Array Ultrasound Probe},
  shorttitle = {Looking beyond the Imaging Plane},
  author = {Xia, Wenfeng and West, Simeon J. and Finlay, Malcolm C. and Mari, Jean-Martial and Ourselin, Sebastien and David, Anna L. and Desjardins, Adrien E.},
  year = {2017},
  month = jun,
  journal = {Scientific Reports},
  volume = {7},
  number = {1},
  pages = {3674},
  issn = {2045-2322},
  doi = {10.1038/s41598-017-03886-4},
  urldate = {2024-11-29},
  langid = {english}
}

@incollection{xiaUltrasonicNeedleTracking2017,
  title = {Ultrasonic {{Needle Tracking}} with a {{Fibre-Optic Ultrasound Transmitter}} for {{Guidance}} of {{Minimally Invasive Fetal Surgery}}},
  booktitle = {Medical {{Image Computing}} and {{Computer-Assisted Intervention}} - {{MICCAI}} 2017},
  author = {Xia, Wenfeng and Noimark, Sacha and Ourselin, Sebastien and West, Simeon J. and Finlay, Malcolm C. and David, Anna L. and Desjardins, Adrien E.},
  editor = {Descoteaux, Maxime and {Maier-Hein}, Lena and Franz, Alfred and Jannin, Pierre and Collins, D. Louis and Duchesne, Simon},
  year = {2017},
  volume = {10434},
  pages = {637--645},
  publisher = {Springer International Publishing},
  address = {Cham},
  doi = {10.1007/978-3-319-66185-8_72},
  urldate = {2024-11-29},
  isbn = {978-3-319-66184-1 978-3-319-66185-8},
  langid = {english}
}

@article{zengFocalLiverLesions2024,
  title = {Focal Liver Lesions: Multiparametric Microvasculature Characterization via Super-Resolution Ultrasound Imaging},
  shorttitle = {Focal Liver Lesions},
  author = {Zeng, Qian-Qian and An, Shi-Zhe and Chen, Chao-Nan and Wang, Zhen and Liu, Jia-Cheng and Wan, Ming-Xi and Zong, Yu-Jin and Jian, Xiao-Hua and Yu, Jie and Liang, Ping},
  year = {2024},
  month = dec,
  journal = {European Radiology Experimental},
  volume = {8},
  number = {1},
  pages = {138},
  issn = {2509-9280},
  doi = {10.1186/s41747-024-00540-3},
  urldate = {2025-04-23},
  langid = {english}
}

@article{zhaoElectromagneticTrackingNeedle2019,
  title = {An Electromagnetic Tracking Needle Clip: An Enabling Design for Low-Cost Image-Guided Therapy},
  shorttitle = {An Electromagnetic Tracking Needle Clip},
  author = {Zhao, Zhuo and Tse, Zion Tsz Ho},
  year = {2019},
  month = may,
  journal = {Minimally Invasive Therapy \& Allied Technologies},
  volume = {28},
  number = {3},
  pages = {165--171},
  issn = {1364-5706, 1365-2931},
  doi = {10.1080/13645706.2018.1496939},
  urldate = {2024-11-29},
  langid = {english}
}

@inproceedings{47008,
	title	= {Compiling machine learning programs via high-level tracing},
	booktitle = {SysML Conference},
	author	= {Roy Frostig and Matthew Johnson and Chris Leary},
	address = {Stanford, CA, USA},
	month = feb,
	year	= {2018},
	URL	= {https://mlsys.org/Conferences/doc/2018/146.pdf}
	}

@article{allesAdaptiveLightModulation2016,
  title = {Adaptive {{Light Modulation}} for {{Improved Resolution}} and {{Efficiency}} in {{All-Optical Pulse-Echo Ultrasound}}},
  author = {Alles, Erwin J. and Colchester, Richard J. and Desjardins, Adrien E.},
  year = 2016,
  month = jan,
  journal = {IEEE Trans. Ultrason., Ferroelect., Freq. Contr.},
  volume = {63},
  number = {1},
  pages = {83--90},
  issn = {0885-3010},
  doi = {10.1109/TUFFC.2015.2497465},
  urldate = {2025-10-21},
  copyright = {https://ieeexplore.ieee.org/Xplorehelp/downloads/license-information/IEEE.html}
}

@article{arjasNeuralNetworkKalman2022,
  title = {Neural {{Network Kalman Filtering}} for 3-{{D Object Tracking From Linear Array Ultrasound Data}}},
  author = {Arjas, Arttu and Alles, Erwin J. and Maneas, Efthymios and Arridge, Simon and Desjardins, Adrien and Sillanpaa, Mikko J. and Hauptmann, Andreas},
  year = 2022,
  month = may,
  journal = {IEEE Trans. Ultrason., Ferroelect., Freq. Contr.},
  volume = {69},
  number = {5},
  pages = {1691--1702},
  issn = {0885-3010, 1525-8955},
  doi = {10.1109/TUFFC.2022.3162097},
  urldate = {2025-10-22},
  copyright = {https://creativecommons.org/licenses/by/4.0/legalcode}
}

@book{bishopPatternRecognitionMachine2006,
  title = {Pattern Recognition and Machine Learning},
  author = {Bishop, Christopher M.},
  year = 2006,
  series = {Information Science and Statistics},
  publisher = {Springer},
  address = {New York},
  isbn = {978-0-387-31073-2},
  langid = {english},
  lccn = {006.4}
}

@article{byrdLimitedMemoryAlgorithm1995,
  title = {A {{Limited Memory Algorithm}} for {{Bound Constrained Optimization}}},
  author = {Byrd, Richard H. and Lu, Peihuang and Nocedal, Jorge and Zhu, Ciyou},
  year = 1995,
  month = sep,
  journal = {SIAM J. Sci. Comput.},
  volume = {16},
  number = {5},
  pages = {1190--1208},
  publisher = {Society for Industrial \& Applied Mathematics (SIAM)},
  issn = {1064-8275, 1095-7197},
  doi = {10.1137/0916069},
  urldate = {2025-07-09},
  langid = {english}
}

@article{choiRelationshipRectusAbdominis2017,
  title = {Relationship between Rectus Abdominis Muscle Thickness and Metabolic Syndrome in Middle-Aged Men},
  author = {Choi, Eun Sil and Cho, Soo Hyun and Kim, Jung-Ha},
  editor = {Ito, Etsuro},
  year = 2017,
  month = sep,
  journal = {PLoS ONE},
  volume = {12},
  number = {9},
  pages = {e0185040},
  issn = {1932-6203},
  doi = {10.1371/journal.pone.0185040},
  urldate = {2026-02-03},
  langid = {english}
}

@online{deepsighttechnologyinc.DeepSightTechnologyReceives,
  title = {{{DeepSight}}™ {{Technology Receives First FDA}} 510(k) {{Clearance}} for {{NeedleVue}}™ {{LC1 Ultrasound System}}},
  author = {DeepSight Technology, Inc.},
  urldate = {2025-10-23},
  url = {https://www.prnewswire.com/news-releases/deepsight-technology-receives-first-fda-510k-clearance-for-needlevue-lc1-ultrasound-system-302529247.html},
  langid = {english}
}

@article{draghiUltrasoundExaminationLiver2007,
  title = {Ultrasound Examination of the Liver: {{Normal}} Vascular Anatomy},
  shorttitle = {Ultrasound Examination of the Liver},
  author = {Draghi, F. and Rapaccini, G.L. and Fachinetti, C. and De Matthaeis, N. and Battaglia, S. and Abbattista, T. and Busilacchi, P.},
  year = 2007,
  month = mar,
  journal = {Journal of Ultrasound},
  volume = {10},
  number = {1},
  pages = {5--11},
  issn = {19713495},
  doi = {10.1016/j.jus.2007.02.002},
  urldate = {2026-02-03},
  copyright = {https://www.elsevier.com/tdm/userlicense/1.0/},
  langid = {english}
}

@book{duckPhysicalPropertiesTissues1990,
  title = {Physical {{Properties}} of {{Tissues}}: {{A Comprehensive Reference Book}}},
  shorttitle = {Physical {{Properties}} of {{Tissues}}},
  author = {Duck, Francis A.},
  year = 1990,
  publisher = {Elsevier Science},
  address = {Burlington},
  isbn = {978-0-12-222800-1},
  langid = {english}
}

@article{kimThicknessRectusAbdominis2012,
  title = {Thickness of {{Rectus Abdominis Muscle}} and {{Abdominal Subcutaneous Fat Tissue}} in {{Adult Women}}: {{Correlation}} with {{Age}}, {{Pregnancy}}, {{Laparotomy}}, and {{Body Mass Index}}},
  shorttitle = {Thickness of {{Rectus Abdominis Muscle}} and {{Abdominal Subcutaneous Fat Tissue}} in {{Adult Women}}},
  author = {Kim, Jungmin and Lim, Hyoseob and Lee, Se Il and Kim, Yu Jin},
  year = 2012,
  month = sep,
  journal = {Arch Plast Surg},
  volume = {39},
  number = {05},
  pages = {528--533},
  issn = {2234-6163, 2234-6171},
  doi = {10.5999/aps.2012.39.5.528},
  urldate = {2026-02-03},
  copyright = {https://creativecommons.org/licenses/by-nc/4.0/},
  langid = {english}
}

@misc{liangVolumetricUltrasoundImaging2025,
  title = {Volumetric Ultrasound Imaging with a Sparse Matrix Array and Integrated Fiber-Optic Sensing for Robust Needle Tracking in Interventional Procedures},
  author = {Liang, Weidong and Rostami, Javad and Baker, Christian and West, Simeon and Diamantopoulos, Athanasios and Mathews, Sunish and Desjardins, Adrien E. and Ourselin, Sebastien and Peralta, Laura and Xia, Wenfeng},
  year = 2025,
  month = sep,
  number = {arXiv:2509.11310},
  eprint = {2509.11310},
  primaryclass = {physics},
  publisher = {arXiv},
  doi = {10.48550/arXiv.2509.11310},
  urldate = {2025-12-01},
  archiveprefix = {arXiv}
}

@article{noimarkPolydimethylsiloxaneCompositesOptical2018,
  title = {Polydimethylsiloxane {{Composites}} for {{Optical Ultrasound Generation}} and {{Multimodality Imaging}}},
  author = {Noimark, Sacha and Colchester, Richard J. and Poduval, Radhika K. and Maneas, Efthymios and Alles, Erwin J. and Zhao, Tianrui and Zhang, Edward Z. and Ashworth, Michael and Tsolaki, Elena and Chester, Adrian H. and Latif, Najma and Bertazzo, Sergio and David, Anna L. and Ourselin, Sebastien and Beard, Paul C. and Parkin, Ivan P. and Papakonstantinou, Ioannis and Desjardins, Adrien E.},
  year = 2018,
  month = feb,
  journal = {Adv Funct Materials},
  volume = {28},
  number = {9},
  pages = {1704919},
  issn = {1616-301X, 1616-3028},
  doi = {10.1002/adfm.201704919},
  urldate = {2025-10-22},
  copyright = {http://onlinelibrary.wiley.com/termsAndConditions\#vor},
  langid = {english}
}

@article{shiLearningbasedSoundSpeed2024,
  title = {Learning-Based Sound Speed Estimation and Aberration Correction for Linear-Array Photoacoustic Imaging},
  author = {Shi, Mengjie and Vercauteren, Tom and Xia, Wenfeng},
  year = 2024,
  month = aug,
  journal = {Photoacoustics},
  volume = {38},
  pages = {100621},
  issn = {22135979},
  doi = {10.1016/j.pacs.2024.100621},
  urldate = {2026-02-09},
  langid = {english}
}

@inproceedings{shishitaniChangeAcousticImpedance2012,
  title = {Change in Acoustic Impedance and Sound Speed of Excised Chicken Breast Muscle Bu High-Intensity Focussed Ultrasound ({{HIFU}}) Exposure},
  booktitle = {Nano-{{Biomedical Engineering}} 2012},
  author = {Shishitani, Takashi and Yoshizawa, Shin and Umemura, Shin-Ichiro},
  year = 2012,
  month = feb,
  pages = {317--324},
  publisher = {Imperial College Press},
  address = {Sakura Hall, Tohoku University, Sendai Japan},
  doi = {10.1142/9781848169067_0037},
  urldate = {2025-09-09},
  isbn = {978-1-84816-905-0 978-1-84816-906-7},
  langid = {english}
}

@article{storchleMeasurementMeanSubcutaneous2018,
  title = {Measurement of Mean Subcutaneous Fat Thickness: Eight Standardised Ultrasound Sites Compared to 216 Randomly Selected Sites},
  shorttitle = {Measurement of Mean Subcutaneous Fat Thickness},
  author = {St{\"o}rchle, Paul and M{\"u}ller, Wolfram and Sengeis, Marietta and Lackner, Sonja and Holasek, Sandra and {F{\"u}rhapter-Rieger}, Alfred},
  year = 2018,
  month = nov,
  journal = {Sci Rep},
  volume = {8},
  number = {1},
  pages = {16268},
  issn = {2045-2322},
  doi = {10.1038/s41598-018-34213-0},
  urldate = {2026-02-03},
  langid = {english}
}

@article{tanakaSplitbasedElevationalLocalization2024,
  title = {Split-Based Elevational Localization of Photoacoustic Guidewire Tip by {{1D}} Array Probe Using Spatial Impulse Response},
  author = {Tanaka, Tomohiko and Imai, Ryo and Takeshima, Hirozumi},
  year = 2024,
  month = mar,
  journal = {Phys. Med. Biol.},
  volume = {69},
  number = {6},
  pages = {065013},
  issn = {0031-9155, 1361-6560},
  doi = {10.1088/1361-6560/ad27fe},
  urldate = {2025-11-19}
}

@article{treebyKWaveMATLABToolbox2010,
  title = {K-{{Wave}}: {{MATLAB}} Toolbox for the Simulation and Reconstruction of Photoacoustic Wave Fields},
  shorttitle = {K-{{Wave}}},
  author = {Treeby, Bradley E. and Cox, B. T.},
  year = 2010,
  journal = {J. Biomed. Opt.},
  volume = {15},
  number = {2},
  pages = {021314},
  issn = {10833668},
  doi = {10.1117/1.3360308},
  urldate = {2026-01-28},
  langid = {english}
}

@article{virtanenSciPy10Fundamental2020,
  title = {{{SciPy}} 1.0: Fundamental Algorithms for Scientific Computing in {{Python}}},
  shorttitle = {{{SciPy}} 1.0},
  author = {Virtanen, Pauli and Gommers, Ralf and Oliphant, Travis E. and Haberland, Matt and Reddy, Tyler and Cournapeau, David and Burovski, Evgeni and Peterson, Pearu and Weckesser, Warren and Bright, Jonathan and Van Der Walt, St{\'e}fan J. and Brett, Matthew and Wilson, Joshua and Millman, K. Jarrod and Mayorov, Nikolay and Nelson, Andrew R. J. and Jones, Eric and Kern, Robert and Larson, Eric and Carey, C J and Polat, {\.I}lhan and Feng, Yu and Moore, Eric W. and VanderPlas, Jake and Laxalde, Denis and Perktold, Josef and Cimrman, Robert and Henriksen, Ian and Quintero, E. A. and Harris, Charles R. and Archibald, Anne M. and Ribeiro, Ant{\^o}nio H. and Pedregosa, Fabian and Van Mulbregt, Paul and {SciPy 1.0 Contributors} and Vijaykumar, Aditya and Bardelli, Alessandro Pietro and Rothberg, Alex and Hilboll, Andreas and Kloeckner, Andreas and Scopatz, Anthony and Lee, Antony and Rokem, Ariel and Woods, C. Nathan and Fulton, Chad and Masson, Charles and H{\"a}ggstr{\"o}m, Christian and Fitzgerald, Clark and Nicholson, David A. and Hagen, David R. and Pasechnik, Dmitrii V. and Olivetti, Emanuele and Martin, Eric and Wieser, Eric and Silva, Fabrice and Lenders, Felix and Wilhelm, Florian and Young, G. and Price, Gavin A. and Ingold, Gert-Ludwig and Allen, Gregory E. and Lee, Gregory R. and Audren, Herv{\'e} and Probst, Irvin and Dietrich, J{\"o}rg P. and Silterra, Jacob and Webber, James T and Slavi{\v c}, Janko and Nothman, Joel and Buchner, Johannes and Kulick, Johannes and Sch{\"o}nberger, Johannes L. and De Miranda Cardoso, Jos{\'e} Vin{\'i}cius and Reimer, Joscha and Harrington, Joseph and Rodr{\'i}guez, Juan Luis Cano and {Nunez-Iglesias}, Juan and Kuczynski, Justin and Tritz, Kevin and Thoma, Martin and Newville, Matthew and K{\"u}mmerer, Matthias and Bolingbroke, Maximilian and Tartre, Michael and Pak, Mikhail and Smith, Nathaniel J. and Nowaczyk, Nikolai and Shebanov, Nikolay and Pavlyk, Oleksandr and Brodtkorb, Per A. and Lee, Perry and McGibbon, Robert T. and Feldbauer, Roman and Lewis, Sam and Tygier, Sam and Sievert, Scott and Vigna, Sebastiano and Peterson, Stefan and More, Surhud and Pudlik, Tadeusz and Oshima, Takuya and Pingel, Thomas J. and Robitaille, Thomas P. and Spura, Thomas and Jones, Thouis R. and Cera, Tim and Leslie, Tim and Zito, Tiziano and Krauss, Tom and Upadhyay, Utkarsh and Halchenko, Yaroslav O. and {V{\'a}zquez-Baeza}, Yoshiki},
  year = 2020,
  month = mar,
  journal = {Nat Methods},
  volume = {17},
  number = {3},
  pages = {261--272},
  publisher = {{Springer Science and Business Media LLC}},
  issn = {1548-7091, 1548-7105},
  doi = {10.1038/s41592-019-0686-2},
  urldate = {2025-07-09},
  copyright = {https://creativecommons.org/licenses/by/4.0},
  langid = {english}
}

@article{wangUltrasonicSoundSpeed2023a,
  title = {Ultrasonic {{Sound Speed Estimation}} for {{Liver Fat Quantification}}: {{A Review}} by the {{AIUM-RSNA QIBA Pulse-Echo Quantitative Ultrasound Initiative}}},
  shorttitle = {Ultrasonic {{Sound Speed Estimation}} for {{Liver Fat Quantification}}},
  author = {Wang, Xiaohong and Bamber, Jeffrey C. and {Esquivel-Sirvent}, Raul and Ormachea, Juvenal and Sidhu, Paul S. and Thomenius, Kai E. and Schoen, Scott and Rosenzweig, Stephen and Pierce, Theodore T.},
  year = 2023,
  month = nov,
  journal = {Ultrasound in Medicine \& Biology},
  volume = {49},
  number = {11},
  pages = {2327--2335},
  issn = {03015629},
  doi = {10.1016/j.ultrasmedbio.2023.06.021},
  urldate = {2026-02-03},
  langid = {english}
}

\end{document}